# Selective reflection of light in glassforming ternary liquid crystalline mixtures

Aleksandra Deptuch[1,*], Zuzanna Zając[2], Marcin Piwowarczyk[1], Anna Drzewicz[1], Marcin Kozieł[3], Magdalena Urbańska[4], Ewa Juszyńska-Gałązka[1,5]

[1] Institute of Nuclear Physics, Polish Academy of Sciences, Radzikowskiego 152, PL-31342 Kraków, Poland

[2] Faculty of Materials Science and Ceramics, AGH University of Cracow, PL-30059 Kraków, Mickiewicza 30, Poland

[3] Faculty of Chemistry, Jagiellonian University, Gronostajowa 2, PL-30387, Kraków, Poland

[4] Institute of Chemistry, Military University of Technology, Kaliskiego 2, PL-00908 Warsaw, Poland

[5] Research Center for Thermal and Entropic Science, Graduate School of Science, Osaka University, 560-0043 Osaka, Japan

*corresponding author, aleksandra.deptuch@ifj.edu.pl

## Abstract

Two ternary liquid crystalline mixtures are formulated and investigated by differential scanning calorimetry, polarizing optical microscopy, X-ray diffraction, and broadband dielectric spectroscopy. Paraelectric smectic A*, ferroelectric smectic C*, and antiferroelectric smectic $C_A$* phases are detected. The glass of the smectic $C_A$* phase is formed at moderate cooling rates. Vitrification prevents the approaching transition to a hexatic smectic phase. One mixture shows a strong thermochromic effect in the smectic $C_A$* phase and selectively reflects blue light in the glassy state. Both mixtures reflect either green or red light in the smectic C* phase, depending on temperature treatment: whether the sample is cooled or heated, or at which rate the temperature changes.

## 1. Introduction

Chiral nematic (cholesteric) and chiral tilted smectic phases (ferroelectric smectic C* and antiferroelectric smectic $C_A$*) are characterized by a helical structure, with a helix pitch often falling in the 100-1000 nm range. This enables selective reflection of light (SRL) from the ultraviolet, visible, and infra-red range by properly aligned samples, with a helix axis perpendicular to the sample's plane: planarly aligned cholesterics (molecules oriented parallel to the sample's plane) and homeotropically aligned SmC* and $SmC_A$* phases (smectic layers parallel to the sample's plane) [1-6]. The wavenumber of reflected light $\lambda_r$ depends on an average refractive index $n_{av}$ and a helix pitch $p$ in a liquid crystal as $\lambda_r = n_{av}p$ [4-7]. The helix pitch can be tuned by temperature [2-4], electric field [6,7], and ultraviolet irradiation (if the photosensitive dopants are present) [6]. The change of color upon cooling or heating is called thermochromism [2]. Liquid crystals exhibiting SRL have a potential use in thermometers, temperature mapping, and optical filtering [2,8,9].

This study aimed to formulate liquid crystalline mixtures forming the SmC* and $SmC_A$* phases and showing SRL, possibly with a thermochromic effect. The components of mixtures are shown in Figure 1. The (*S*)-4-[(1-methylheptyloxy)carbonyl]phenyl 4′-octyloxy-4-biphenylcarboxylate (MHPOBC) shows a strong dependence of a helix pitch on temperature in the $SmC_A$* phase, including a helix inversion, which means that a helix pitch initially increases with temperature until unwinding, which is followed by a decrease in a helix pitch with temperature [3]. As such, MHPOBC is a proper base component for a thermochromic mixture. Another selected components are (*S*)-4′-(1-methylheptyloxycarbonyl)biphenyl-4-yl 4-[5-(2,2,3,3,4,4,4-heptafluorobutoxy)pentyl-1-oxy]-2-fluorobenzoate (3F5HPhF6), (*S*)-4′-(1-methylheptyloxycarbonyl)biphenyl-4-yl 4-[6-(2,2,3,3,4,4,4-heptafluorobutoxy)hexyl-1-oxy]-2-fluorobenzoate (3F6HPhF6), (*S*)-4’-(1-methylheptylcarbonyl)biphenyl-4-yl 4-[5-(2,2,3,3,4,4,4-heptafluorobutoxy)pentyl-1-oxy]benzoate (3F5HPhH6), and (*S*)-4’-(1-methylheptylcarbonyl)biphenyl-4-yl 4-[6-(2,2,3,3,4,4,4-heptafluorobutoxy)hexyl-1-oxy]benzoate (3F6HPhH6). A helix pitch in the SmC$_A$* phase shows an inversion for 3F5HPhF6, increases on heating for 3F5HPhH6, and is rather constant for 3F6HPhF6 and 3F6HPhH6 [4], therefore, a helix pitch in the main MHPOBC component may be modified to a varying extent. Moreover, 3F5HPhF6, 3F5HPhH6, and 3F6HPhF6 form the smectic glass on cooling at moderate rates [10-12], which makes it easier to obtain a glassforming mixture, where crystallization is hindered below the glass transition temperature and optical properties are preserved [8].

The tested compositions are MHPOBC : 3F5HPhF6 : 3F5HPhH6 with a molar ratio 0.5 : 0.25 : 0.25, denoted as MIX5HFHH6, and MHPOBC : 3F6HPhF6 : 3F6HPhH6 with a molar ratio 0.5 : 0.25 : 0.25, denoted as MIX6HFHH6. In the first part, the differential scanning calorimetry (DSC) at various cooling/heating rates, combined with polarizing optical microscopy (POM) observations, determines the phase sequence and glassforming properties of ternary mixtures. The second part presents the observation of SRL in homeotropically aligned samples in the SmC* and SmC$_A$* phases. In the third part, the X-ray diffraction (XRD) is used to confirm the absence of crystallization at room temperature and to determine the layer spacing in smectic phases. In the fourth part, the broadband dielectric spectroscopy (BDS) is applied to investigate relaxation times, which are further used to confirm the identification of smectic phases and the investigation of the glass transition in the SmC$_A$* phase. Specifically, the temperature dependences of the α-relaxation time in mixtures are compared with those for glassforming pure components 3F5HPhF6, 3F5HPhH6, and 3F6HPhF6 [11,12].

**MHPOBC**

$C_8H_{17}O$–Ph–Ph–COO–Ph–COO–$C^*H(CH_3)C_6H_{13}$ (*S*)

**3FmHPhF6, m = 5, 6**

$C_3F_7CH_2OC_mH_{2m}O$–Ph(F)–COO–Ph–Ph–COO–$C^*H(CH_3)C_6H_{13}$ (*S*)

**3FmHPhH6, m = 5, 6**

$C_3F_7CH_2OC_mH_{2m}O$–Ph–COO–Ph–Ph–COO–$C^*H(CH_3)C_6H_{13}$ (*S*)

Figure 1. Molecular structures of MIXmHFHH6 (m = 5, 6) components.

## 2. Experimental details

The ternary mixtures denoted as MIXmHFHH6 (m = 5, 6) have compositions:

- MIX5HFHH6 – MHPOBC : 3F5HPhF6 : 3F5HPhH6 molar ratio 0.4988(5) : 0.2498(4) : 0.2514(4),
- MIX6HFHH6 – MHPOBC : 3F6HPhF6 : 3F6HPhH6 molar ratio 0.5003(5) : 0.2496(3) : 0.2500(4).

All components were synthesized in the Institute of Chemistry of the Military University of Technology according to routes described elsewhere [4,13]. The components of each mixture were dissolved in acetone and mixed in solution. The solution was heated to 318 K for faster evaporation of acetone. The precipitate was heated to 433 K and cooled slowly to room temperature.

The DSC thermograms were collected with the TA Instruments DSC 2500 calorimeter (cooling and heating at 2, 5, 10, 15, 20, 25, 30 K/min in 173-433 K). The MIXmHFHH6 samples, weighting 13.9 mg (m = 5) and 10.04 mg (m = 6), were contained within aluminum pans. Data were analyzed in TRIOS.

The microscopic measurements were performed with the Leica DM2700 P polarizing microscope in the transmission mode for observation of POM textures (samples between two glass slides without aligning layers, cooling and heating at 10 K/min in 188-433 K) and in the reflection mode for observation of SRL (AWAT cells of 5 μm thickness with polymer layer providing a homeotropic alignment, cooling and heating at 10 K/min in 188-423 K and 2 K/min in 383-423 K for m = 5 and 373-423 K for m = 6). Data were analyzed in TOApy [14] and ImageJ [15].

The XRD patterns were collected with the PANalytical X'Pert PRO diffractometer (Bragg-Brentano geometry, 2θ = 2-30°, CuKα radiation, $\lambda_{CuK\alpha1}$ = 1.540562 Å, $\lambda_{CuK\alpha2}$ = 1.544390 Å [16]) at room temperature. The samples were placed in flat sample holders of 13 mm × 10 mm × 0.2 mm size. The 2θ calibration was based on the NIST Standard Reference Material 675 [17], supplied by Merck. The XRD patterns as a function of temperature were collected on cooling in 188-423 K and on heating to 273 K and 298 K using the AntonPaar TTK 450 temperature stage. Data were analyzed in WinPLOTR [18] and OriginPro.

The BDS spectra were collected with the Novocontrol impedance spectrometer (cooling and heating in 173-433 K, frequency 0.1-$10^6$ Hz). The samples with 75 μm thickness were contained between gold electrodes, with an active area of 10 mm, without aligning layers and with polytetrafluoroethylene spacers. Data were analyzed in OriginPro.

## 3. Results

### 3.1. Phase sequence

The DSC thermograms of MIX5HFHH6 (Figure 2) and MIX6HFHH6 (Figure 3), together with POM observations in the transmission mode (Figures S1-S4 in the supplementary materials), indicate three smectic phases: SmA*, SmC*, both present only in a narrow temperature range, and $SmC_A$* with a much wider temperature range. The onset temperatures of minima and maxima in DSC thermograms were taken as phase transition temperatures [19]. The exceptions are the SmC*/$SmC_A$* transition on cooling and the SmC*/SmA* transition on heating in MIX5HFHH6, where onsets were difficult to obtain due to overlapping peaks, and peak temperatures were considered instead [19]. The phase transition temperatures are collected in Table 1 together with corresponding enthalpy changes. The glass transition of the $SmC_A$* phase shows as a step in the heat capacity [20]. The glass transition temperature, obtained as a half-height of this step, is equal to 240(1) K / 242.8(5) K on cooling/heating for MIX5HFHH6 and 233(2) K / 235.3(6) K on cooling/heating for MIX6HFHH6 when extrapolated to 0 K/min rate. MIX5HFHH6 does not crystallize on cooling, and cold crystallization is observed only for lower heating rates: 2-10 K/min. The onset temperature of cold crystallization is 271-276 K. The transition between two crystal phases is observed at 294-303 K and melting to the $SmC_A$* phase has an onset at 302-310 K. MIX6HFHH6 crystallizes partially only on cooling at 2 K/min with an onset at 276 K. The cold crystallization occurs for all heating rates from the 2-30 K/min range. Cold crystallization temperature increases from 260 K at 2 K/min to 281 K at 30 K/min. The crystal phase melts with an onset at 293-301 K.

The POM textures of MIX5HFHH6 reveal a homeotropic alignment of a sample on cooling (Figure S1). The helix inversion shows a double maximum in luminance surrounding a minimum at ca. 360 K. The color change from red to blue is observed below 320 K due to a decreased helix pitch. An appearance of light areas, leading to a step-like increase in luminance at ~260 K is attributed to the glass transition. However, it is at a higher temperature than obtained by DSC. The POM textures of MIX5HFHH6 on heating confirm two crystal phases and a transition between them is visible as a rapid decrease in blue and green components (Figure S2). Crystallization destroys a homeotropic alignment, and a strongly defective texture is observed above the melting temperature. The homeotropic alignment appears back spontaneously in the SmA* phase.

The POM textures of MIX6HFHH6 are fan-like, indicating a planar alignment of the sample on cooling (Figure S3). Only small homeotropically-aligned areas increase in intensity when temperature decreases, indicating a helix inversion. The glass transition does not influence textures on cooling, while on heating the glass softening can be detected by the beginning of cold crystallization (Figure S4). The planar alignment is preserved in the crystal phase, and fan-shaped textures of smectic phases are also observed when heating. At the same time, the previously homeotropically-aligned areas are strongly defective above the melting temperature.

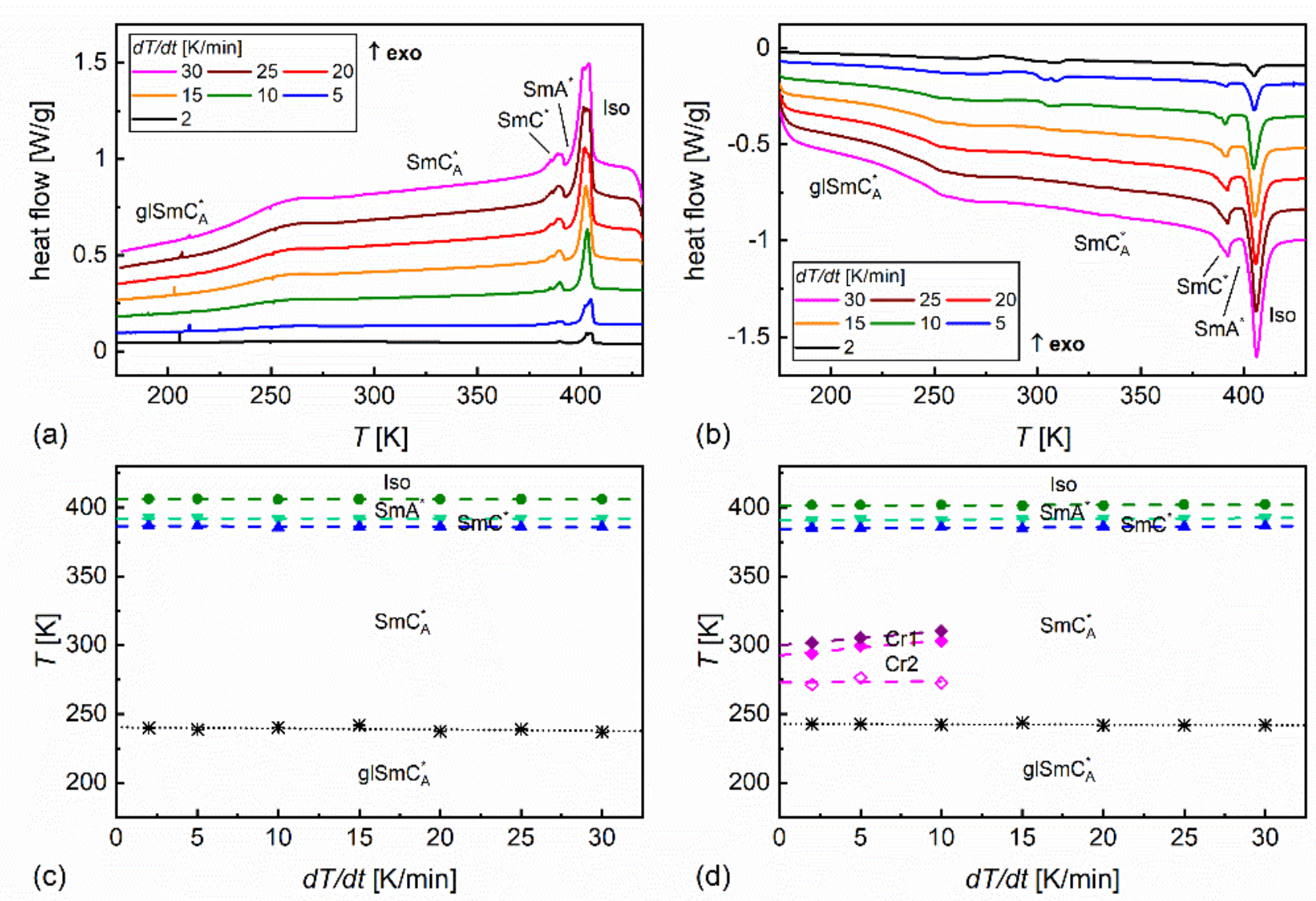

Figure 2. DSC thermograms for MIX5HFHH6 at cooling (a) and heating (b) with corresponding transition temperatures vs. cooling rate (c) and heating rate (d).

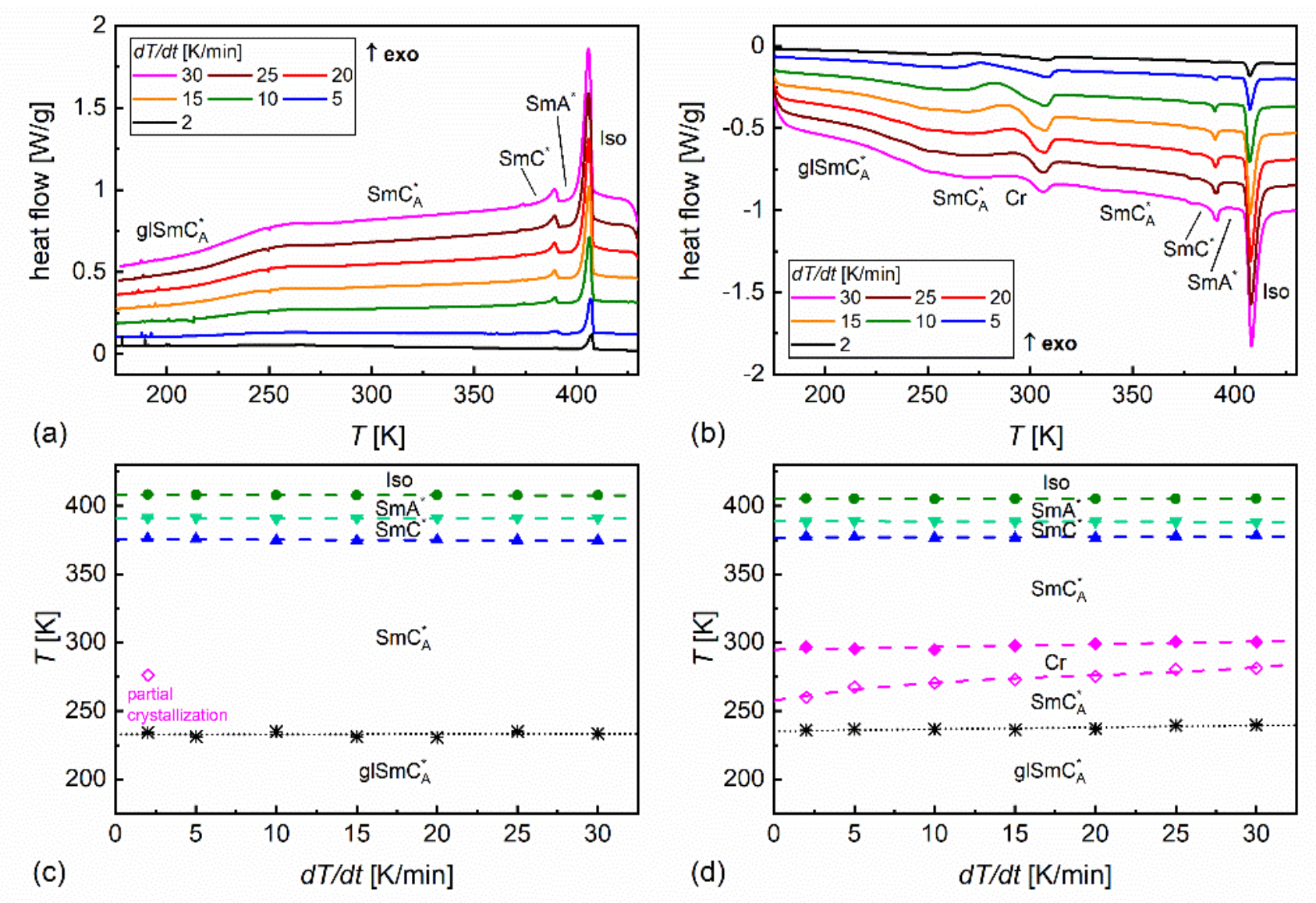

Figure 3. DSC thermograms for MIX6HFHH6 at cooling (a) and heating (b) with corresponding transition temperatures vs. cooling rate (c) and heating rate (d).

Table 1. Transitions between the smectic phases and isotropic liquid in MIXmHFHH6 (m = 5, 6) investigated by DSC on cooling/heating: onset temperatures in K (1st row) extrapolated to 0 K/min and enthalpy change in kJ/mol (*2nd row*).

| sample | $SmC_A$*/SmC* | SmC*/SmA* | SmA*/Iso |
|---|---|---|---|
| MIX5HFHH6 | 386.5 / 384.6 | 392.2 / 390.7 | 406.2 / 401.6 |
| | *0.2 / 0.2* | *0.5 / 0.6* | *5.1 / 5.2* |
| MIX6HFHH6 | 375.6 / 376.7 | 391.0 / 388.7 | 408.1 / 405.1 |
| | *> 0.1 / > 0.1* | *0.5 / 0.4* | *5.3 / 5.5* |

### 3.2. Selective reflection of visible light

The microscopic images obtained in a reflection mode for MIXmHFHH6 with m = 5, 6 at 10 K/min in 188-423 K are presented in Figures S5-S8, and images obtained at 2 K/min in the high-temperature range around the SmC* phase are presented in Figures S9-S12 in the supplementary materials. Plots of each image's red, green, blue components, and total luminance accompany the figures.

MIX5HFHH6 reflects green light on cooling in the SmC* phase (Figure S5). A maximum in luminance observed in the $SmC_A$* phase at 338 K is interpreted as a helix inversion. On further cooling, SRL in a visible range is observed, with a wavelength decreasing with decreasing temperature, which supports the presence of a helix inversion at 338 K. The maxima of red, green, and blue components occur at 300 K, 293 K, and 285 K, as shown in Figure 4a-c. A dark blue image is observed in a wide 188-270 K range, with an exemplary image at 200 K shown in Figure 4d. It indicates that a helix pitch does not change in the glassy $SmC_A$* phase because SRL does not enter the ultraviolet range, but remains at the brink of the visible range. MIX5HFHH6 undergoes cold crystallization on heating (Figure S6), which misaligns the sample. The maxima of blue, green, and red components shift to 315 K, 322 K, and 330 K, and a helix inversion shifts to 383 K in the $SmC_A$* phase, while in the SmC* phase there is a reflection of red light on heating.

MIX6HFHH6 reflects red light on cooling in the SmC* phase (Figure S7). A helix inversion occurs at 310 K. The thermochromic effect is not observed; SRL occurs likely in the infra-red range. However, in the glassy state, it is approaching the visible range, as some weakly red spots appear in microscopic images. A helix inversion is not observed when heating because of the ongoing melting of the crystal phase (Figure S8). The reflection of red light occurs in the SmC* phase when heating.

The measurements at 2 K/min were performed above the melting temperature; thus, the misalignment of the sample caused by cold crystallization does not occur. MIX5HFHH6 reflects mainly green light when cooled (blue image is observed only just below the SmA*/SmC* transition) and red light when heated at 2 K/min in the SmC* phase (Figure 5a,b), similarly as it was observed for the 10 K/min rate. It suggests that a helix pitch in the SmC* phase is shorter after the SmA*/SmC* transition on cooling than after the $SmC_A$*/SmC* transition on heating. The changes in the sample's alignment, which would affect the refraction index, are not likely because there was no crystallization between cooling and heating runs. MIX6HFHH6 reflects mainly green light with some red contribution on cooling and red light on heating at 2 K/min (Figure 5c,d), contrary to the results for 10 K/min, when reflection of red light only was observed both on cooling and heating. It indicates that for MIX6HFHH6, the rate of temperature change also influences the helix pitch.

The influence of the temperature program on the helix pitch in SmC* was reported in [7], but the observed effect was the opposite: longer helix pitch was obtained for cooling and shorter for heating. It was explained by surface anchoring, i.e., interactions between a liquid crystal and its container. Notably, the sample in [7] was aligned planarly; thus, the surface anchoring may have impacted a helix pitch differently than in homeotropically aligned samples used in this study.

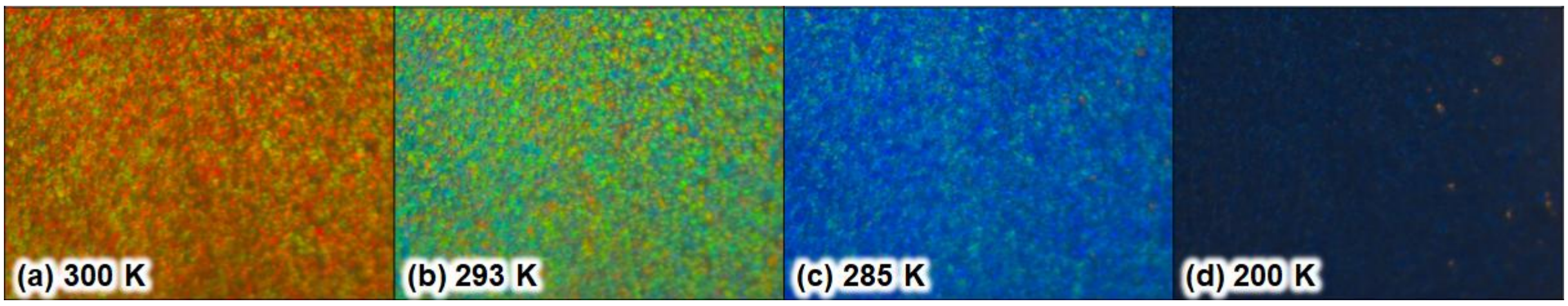

Figure 4. SRL in the SmC$_A$* phase (a, b, c) and glassy SmC$_A$* phase (d) of MIX5HFHH6 observed on cooling at 10 K/min.

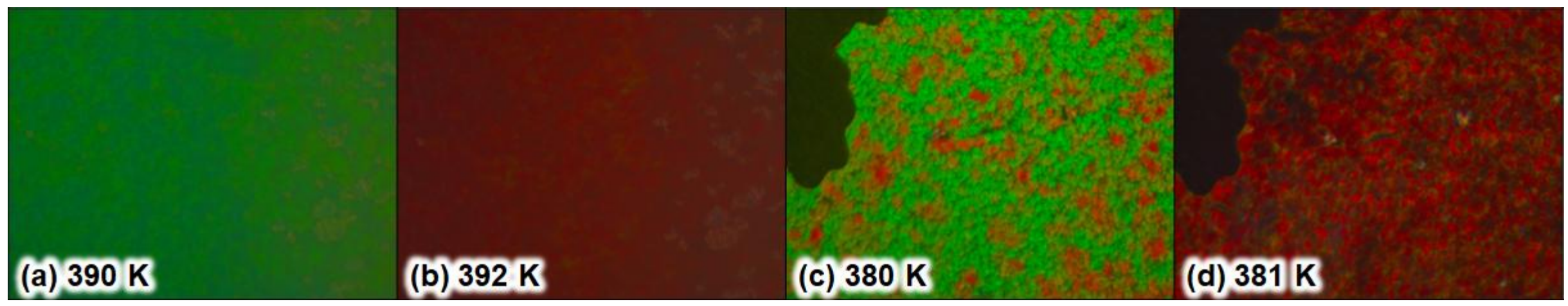

Figure 5. SRL in the SmC* phase of MIX5HFHH6 on cooling (a) and heating (b), and in the SmC* phase of MIX6HFHH6 on cooling (c) and heating (d) at 2 K/min.

### 3.3. Structure of smectic phases

The XRD patterns (Figure 6) were collected at room temperature after cooling samples from the isotropic liquid phase. The alignment of such samples is homeotropic, which is indicated by the very low intensity of the diffuse maximum from intra-layer short-range order at higher angles compared to the sharp peaks from smectic layers at low angles [21]. The homeotropic alignment is also visible as a reflection of green light at room temperature when the MIX5HFHH6 sample is viewed from the top. The same sample looks blue when viewed at an oblique angle at room temperature, which shows an influence of a viewing angle (Figure S13 in the supplementary materials).

The low-angle peaks have Miller indices ($00l$) where $l$ is an integer. Their positions $2\theta_l$ in XRD patterns are related to the smectic layer spacing $d$ by the Bragg formula $l\lambda = 2d\sin(\theta_l - \theta_s)$ [21]. Any systematic shift $\theta_s$ is corrected by considering the positions of all visible ($00l$) peaks (Figure S14). The $d$ values at the room temperature are 32.45(7) Å in MIX5HFHH6 and 33.54(7) Å in MIX6HFHH6. The integrated intensities $I_{00l}$ of ($00l$) peaks depend on the electron density profile along the direction perpendicular to smectic layers [22]:

$$\rho(z) - \rho_0 \propto \sum_{l=1}^{4} \pm |F_{00l}| \cos(2\pi l z/d), \quad (1)$$

where the structure factors $F_{00l}$ are related to integrated intensities $I_{00l}$ and Lorentz-polarization correction $Lp$ as $|F_{00l}| = \sqrt{I_{00l}/Lp}$ [23]. The $F_{00l}$ factors for the centrosymmetric $\rho(z)$ profile are real numbers with unknown ± signs [22,23]. The $Lp$ correction for single crystal samples was applied [23], due to a homeotropic alignment of samples:

$$Lp \propto \frac{1+\cos^2(2\theta)}{\sin(2\theta)}. \quad (2)$$

The minima in $\rho(z)$ are assumed at the borders of smectic layers, namely at $z = 0$ and $z = d$. Therefore, the (−) sign of the main $\cos(2\pi z/d)$ component from Equation (1) has to be selected. After testing different signs of other $\cos(2\pi lz/d)$ components with $l > 1$, the (−) signs for all $F_{00l}$ factors were selected. The obtained $\rho(z)$ distributions are shown in the inset in Figure 6. The main maximum of $\rho(z)$ in the middle of the smectic layer correspond to aromatic cores, while the lateral maxima correspond to the fluorinated terminal chains in 3FmHPhH6 and 3FmHPhF6 molecules. Presence of higher harmonics in XRD patterns indicate that smectic layers are well defined in the $SmC_A$* phase.

The smectic layer spacing, obtained from XRD patterns collected on cooling (Figures S14 and S15), is presented in Figure 7. The layer shrinkage between the largest $d$ in SmA* and the lowest $d$ in $SmC_A$* is equal to 10.4% for m = 5 and 9.5% for m = 6. Next, the layer spacing increases by 3.8% for m = 5 and 5.1% for m = 6 on cooling from 330 K until a maximum at 248 K. The temperature dependence of $d$ in 248-330 K suggests an approaching transition to a more ordered, hexatic smectic $X_A$* phase [24,25]. However, the DSC results exclude the $SmC_A$* → $SmX_A$* transition. The hypothetical $SmC_A$* → $SmX_A$* transition temperature is apparently lower than the glass transition temperature, which prevents formation of the $SmX_A$* phase in MIXmHFHH6. The layer spacing decreases slowly below 248 K, in the $SmC_A$* glass. A local maximum in $d$ in the glass transition region was also observed for pure 3FmHPhF6 components, forming the $SmC_A$* glass [12], but the relative change in $d$ (<1%) was much smaller than for MIXmHFHH6. It supports an assumption of the interrupted $SmC_A$* → $SmX_A$* transition in MIXmHFHH6.

The well-developed crystal phase, with sharp diffraction peaks, is not observed. MIX5HFHH6 shows very weak signs of cold crystallization after heating to 298 K (Figure S14). MIX6HFHH6 crystallizes partially already after gradual cooling to 268 K, but the low-angle peak from the $SmC_A$* phase has still a high intensity, thus, most of the sample remains in the liquid crystal state (Figure S15).

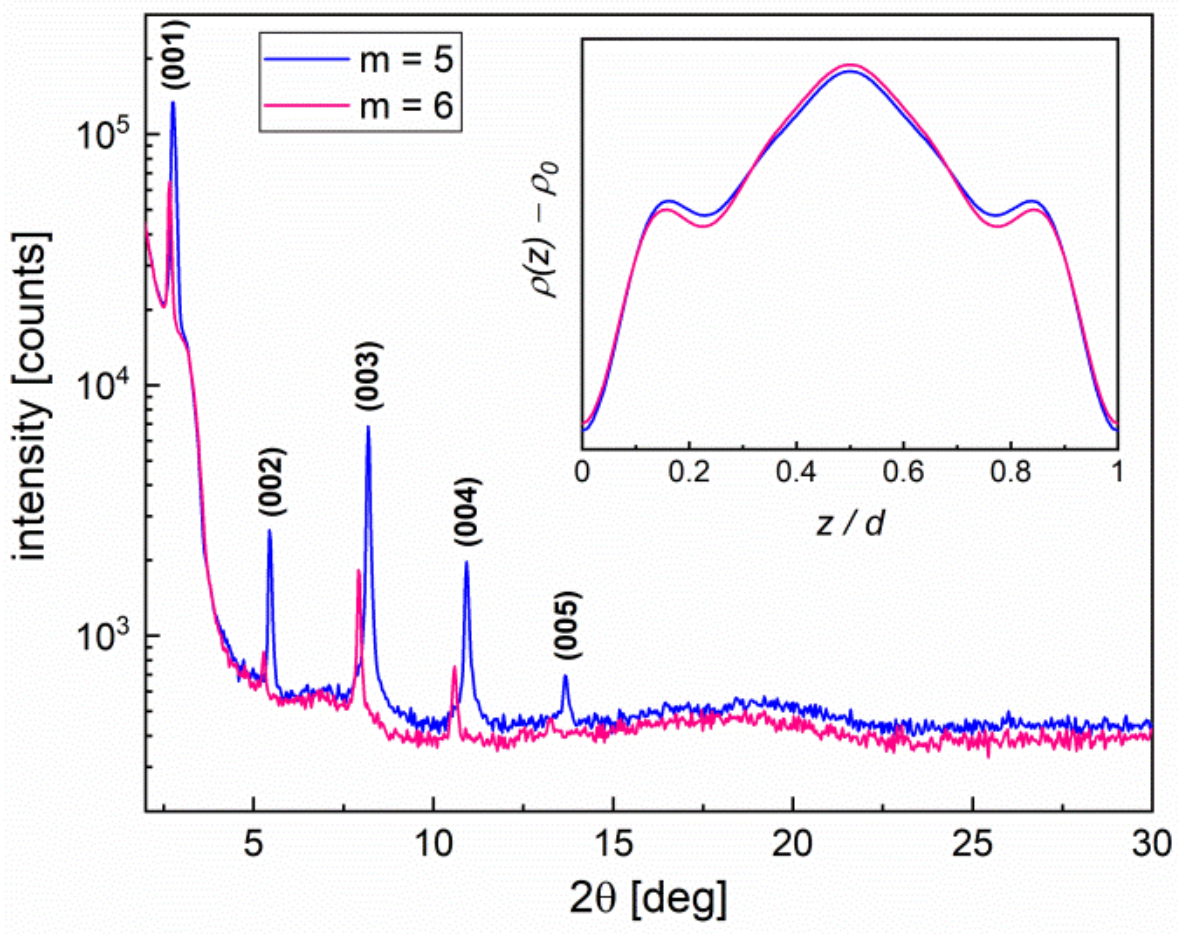

Figure 6. Diffraction patterns of MIXmHFHH6 and corresponding electron density profile along the ⟨001⟩ direction (inset) in the $SmC_A$* phase at room temperature.

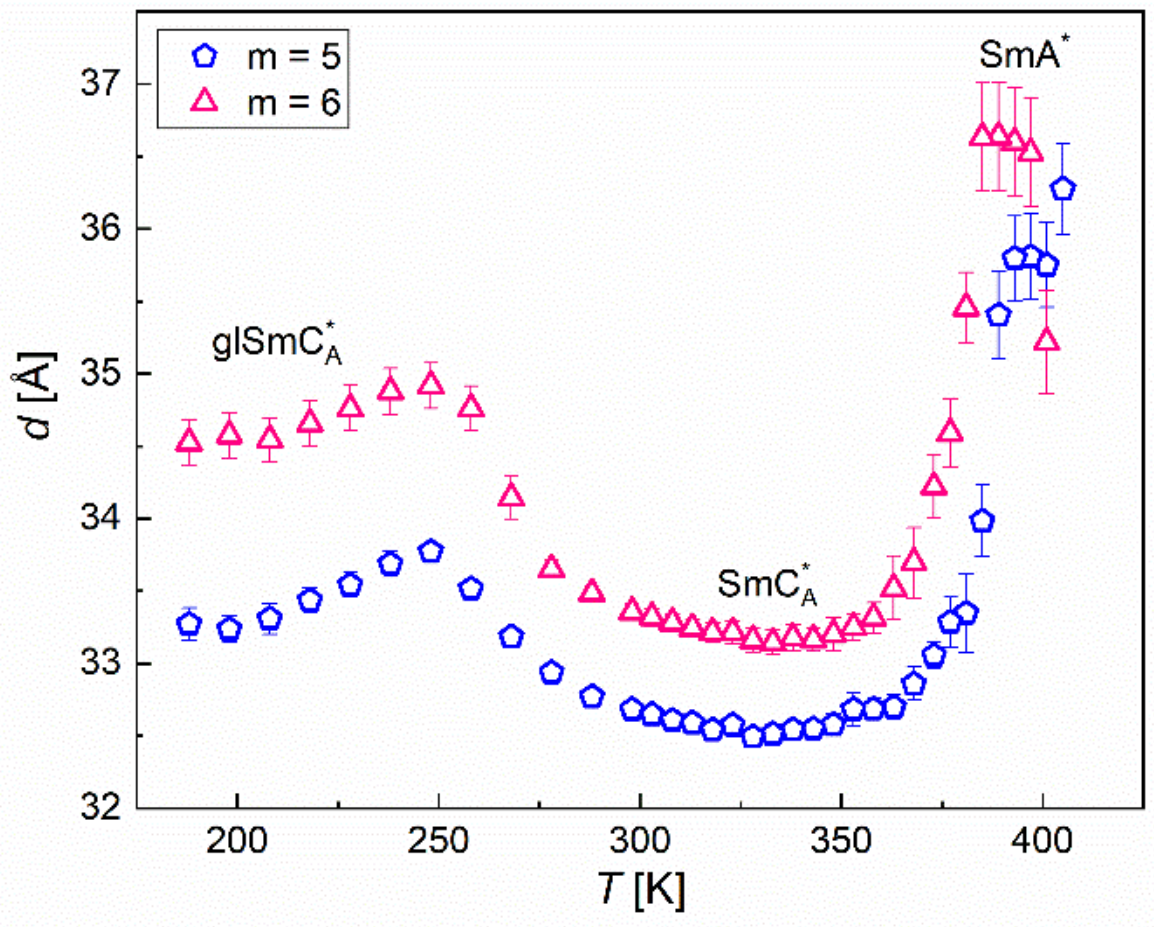

Figure 7. Smectic layer spacing in MIXmHFHH6 as a function of temperature on cooling.

### 3.4. Relaxation processes

The Havriliak-Negami model [26] was applied to fit each process in experimental BDS spectra (Figure 8). The parameters of this model are: relaxation time $\tau_{HN}$, dielectric strength $\Delta\varepsilon$, and shape parameters $a$, $b$. The full formula for fitting the complex dielectric permittivity $\varepsilon^*(f)$ as a function of frequency has a form:

$$\varepsilon^*(f) = \varepsilon'(f) - i\varepsilon''(f) = \varepsilon_\infty + \sum_j \frac{\Delta\varepsilon_j}{\left(1+(2\pi i f \tau_{HNj})^{1-a_j}\right)^{b_j}} - \frac{iS_1}{(2\pi f)^{n_1}} + \frac{S_2}{(2\pi f)^{n_2}}, \quad (3)$$

where: $\varepsilon'(f)$ is dielectric dispersion, $\varepsilon''(f)$ is dielectric absorption, $\varepsilon_\infty$ is dielectric dispersion in the limit of high frequency, and $S_1$, $S_2$, $n_1$, $n_2$ contribute to low-frequency background [27]. If $n_1 = 1$, then $S_1 = \sigma/\varepsilon_0$, where $\sigma$ is ionic conductivity and $\varepsilon_0$ is vacuum permittivity [28].

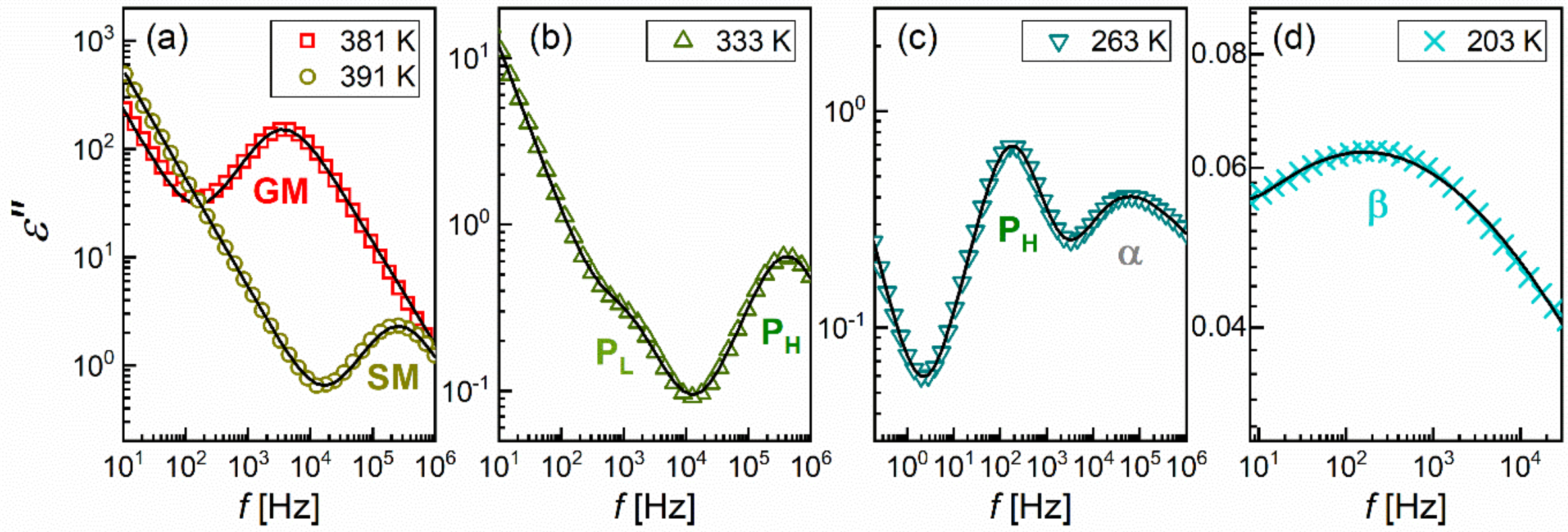

Figure 8. Selected BDS spectra of MIX6HFHH6 (points) in SmA* at 391 K and SmC* at 381 K (a), SmC$_A$* at 333 K (b), SmC$_A$* at 263 K (c), and glassy SmC$_A$* at 203 K (d) with fitting results of Equation (3) (lines).

The relaxation time $\tau$ analyzed in this study (Figure 9) corresponds to the peak position in dielectric absorption. If the shape parameter $b$ = 1, then $\tau = \tau_{HN}$ from Equation (3). However, if $b \neq 1$, then the relaxation time has to be determined as [28]:

$$\tau = \tau_{HN}\left(\sin\left(\frac{\pi(1-a)}{2+2b}\right)\right)^{-\frac{1}{1-a}}\left(\sin\left(\frac{\pi(1-a)b}{2+2b}\right)\right)^{\frac{1}{1-a}}. \quad (4)$$

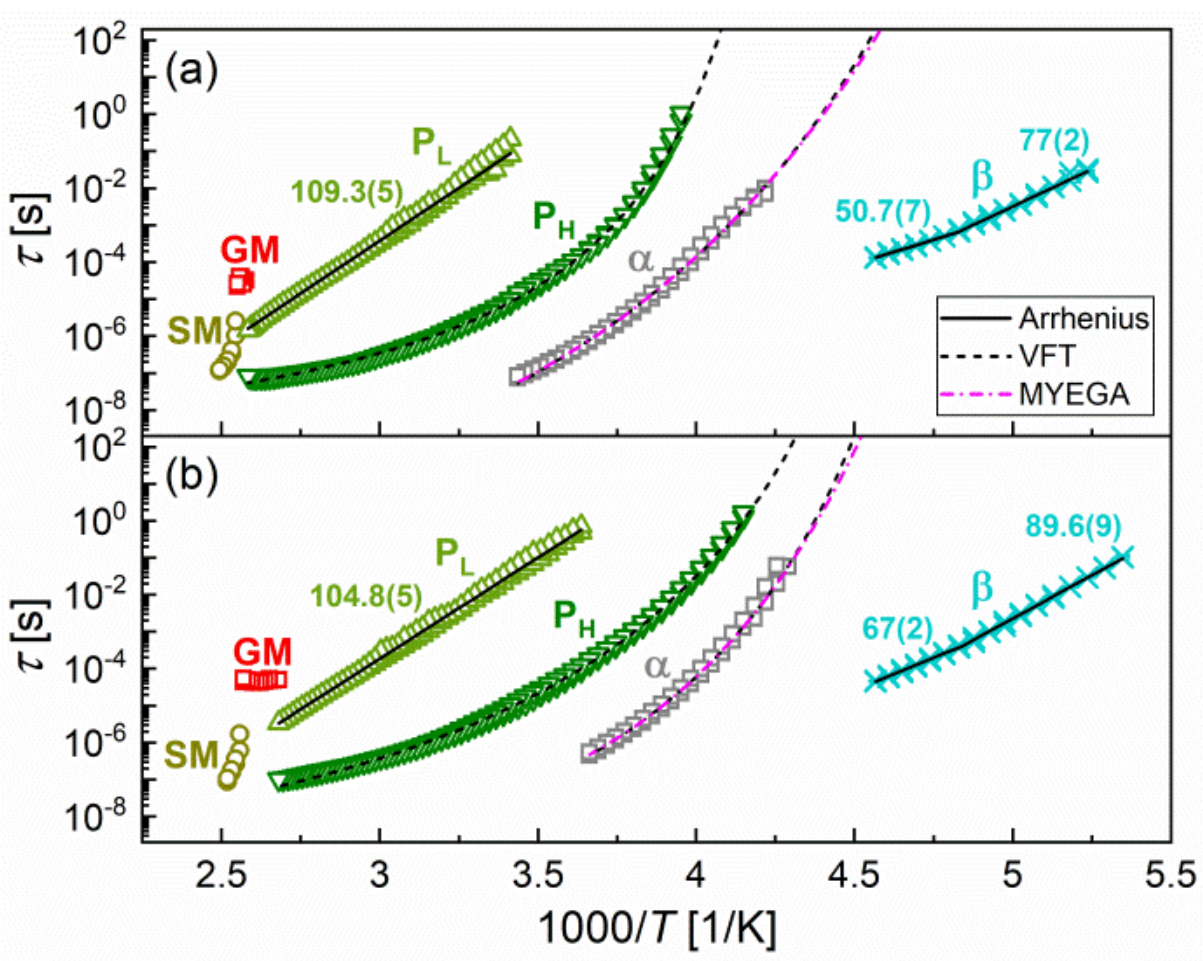

Figure 9. Activation plot of relaxation times obtained from the BDS spectra of MIXmHFHH6 with m = 5 (a) and m = 6 (b). The activation energies are in kJ/mol.

Certain relaxation processes enable the identification of smectic phases. The paraelectric SmA* phase is recognized by the soft mode (SM in Figure 8a, fluctuations of the tilt magnitude) with the relaxation time increasing with decreasing temperature [27]. The ferroelectric SmC* phase is recognized by the Goldstone mode (GM in Figure 8a, fluctuations of the tilt azimuth) with the relaxation time weakly dependent on temperature and the largest dielectric strength [27]. The antiferroelectric SmC$_A$* phase is identified by two phasons at low and high frequencies (P$_L$ and P$_H$ in Figure 8c, in-phase and anti-phase fluctuations of the tilt azimuth in neighbor smectic layers) [29]. The SM, GM, P$_L$, and P$_H$ processes can be fitted with the shape parameter $b = 1$. The P$_L$ process overlaps with the molecular s-process (rotations around molecular short axes) [29] and shows the Arrhenius dependence of the relaxation time on temperature: $\tau(T) = \tau_\infty \exp(E_a/RT)$, where $\tau_\infty$ is a pre-exponential factor and $R$ is the gas constant. The activation energy $E_a$ of P$_L$ is equal to 109.3(5) kJ/mol for MIX5HFHH6 and 104.8(5) kJ/mol for MIX6HFHH6 (Figure 9). The relaxation time of the P$_H$ process deviates from the Arrhenius dependence and is described by the Vogel-Fulcher-Tammann (VFT) formula: $\tau(T) = \tau_\infty \exp(B/(T - T_V))$ [30,31], which is an extended Arrhenius formula with one additional parameter, the Vogel temperature $T_V$ (if $T_V = 0$, then $B = E_a/R$).

Two processes related to the glass transition are observed in BDS spectra of MIX5HFHH6: primary α-relaxation [28,30-32] with $b \neq 1$ (Figure 8c) and very weak secondary β-relaxation [33] with $b = 1$ (Figure 8d; the apparent asymmetry of the absorption peak is caused by overlapping with the tail of a stronger α-relaxation). The α-relaxation can be attributed to rotation around the long molecular axes [32] and β-relaxation to intra-molecular rotations in the case of flexible molecules [33]. The α-relaxation time follows the VFT formula, while the β-relaxation time follows the Arrhenius formula, with an increase in the activation energy below 207 K: $E_a$ = 50.7(7) and 77(2) kJ/mol for MIX5HFHH6, $E_a$ = 67(2) and 89.6(9) kJ/mol for MIX6HFHH6. An increase of $E_a$ at low temperatures suggests that larger parts of molecules are involved in β-relaxation [34].

The glass transition temperature $T_g$ is defined as a temperature where the α-relaxation time is equal to 100 s [30]. The extrapolation of the α-relaxation time to lower temperatures (Figure 9) can be performed by fitting the mentioned earlier VFT formula or alternative Mauro-Yue-Ellison-Gupta-Allan (MYEGA) formula: $\tau(T) = \tau_\infty \exp((K/T)\exp(H/RT))$, where: $H$ is the energy difference between constrained and unconstrained states and $K$ is proportional to the effective activation barrier and inversely proportional to logarithm of number of degenerate configurations [35]. The MYEGA formula is applied rarer than the VFT formula [32], although it has a certain advantage: $\tau \to \infty$ at $T = 0$ according to the MYEGA formula, same as in the Arrhenius dependence, which is more physically reasonable than $\tau \to \infty$ at $T = T_V > 0$ according to the VFT formula. The fitting results of VFT and MYEGA formulas are presented in Tables 2 and 3. For comparison, the results for the glassforming components from previous publications [11,12] are also presented. The α-relaxation times of 3F5HPhH6 analyzed in Ref. [11] were $\tau_{HN}$. In this work, they were calculated to $\tau$ corresponding to peak position in $\varepsilon''$ using Equation (4) for consistency with other data. The MYEGA formula was not applied in [11,12]. The parameters were obtained in this work based on α-relaxation times presented in [11,12]. The $T_g$ values determined from VFT and MYEGA formulas are close for each mixture: 219-220 K for MIX5HFHH6 and 222-223 K for MIX6HFHH6. The $T_g$ values for both mixtures are lower by 8-11 K than these obtained for glassforming components. Also, the Vogel temperature $T_V$ is lower in mixtures than in pure components and the difference is more significant than for $T_g$.

Another parameter characterizing a glassformer is the fragility index $m_f$ defined as:

$$m_f = \left.\frac{\mathrm{d}\log_{10}\tau_\alpha(T)}{\mathrm{d}(T_g/T)}\right|_{T=T_g} \quad (5).$$

The fragility index takes values from the ca. 16-200 range and it increases with increasing deviation of the α-relaxation time from the Arrhenius dependence [30]. The $m_f$ values obtained via the MYEGA model are 6-14% lower than those obtained via the VFT formula. Still, this difference does not change the general conclusion. Both MIXmHFHH6 mixtures and their glassforming components are rather fragile glassformers, with $m_f > 50$ in all cases, but mixtures are less fragile than their components. A lower $m_f$ index is correlated with a lower energy difference $H$ between constrained and unconstrained states and a higher $K$ parameter from the MYEGA model.

Table 2. Fitting parameters of the VFT formula of the α-relaxation time in MIXmHFHH6 mixtures and their glassforming components.

| sample | $\log_{10}(\tau/s)$ | $B$ [K] | $T_V$ [K] | $T_g$ [K] | $m_f$ |
|---|---|---|---|---|---|
| MIX5HFHH6 | −14.3(5) | 2007(151) | 166(3) | 219.9(3) | 67(2) |
| MIX6HFHH6 | −12.0(5) | 1082(117) | 189(3) | 222.9(2) | 93(4) |
| 3F5HPhH6 [11] | −13.1(1) | 1313(29) | 193.5(6) | 231.2(1) | 92.4(8) |
| 3F5HPhF6 [12] | −10.0(1) | 701(13) | 204.0(2) | 229.1(1) | 111.0(4) |
| 3F6HPhF6 [12] | −11.2(4) | 817(56) | 203.1(8) | 230.1(3) | 112(1) |

Table 3. Fitting parameters of the MYEGA formula of the α-relaxation time in MIXmHFHH6 mixtures and their glassforming components.

| sample | $\log_{10}(\tau/s)$ | $H$ [kJ/mol] | $K$ [K] | $T_g$ [K] | $m_f$ |
|---|---|---|---|---|---|
| MIX5HFHH6 | −11.4(3) | 6.5(3) | 190(30) | 219.2(7) | 61.4(5) |
| MIX6HFHH6 | −9.2(3) | 11.4(4) | 12(3) | 222(2) | 80.4(5) |
| 3F5HPhH6 [11] | −10.1(1) | 11.0(1) | 21(2) | 230.4(4) | 81.7(2) |
| 3F5HPhF6 [12] | −7.0(1) | 19.3(1) | 0.18(1) | 228.7(3) | 100.3(5) |
| 3F6HPhF6 [12] | −7.4(2) | 19.3(2) | 0.21(3) | 229.9(8) | 105(2) |

The ionic conductivity $\sigma$, obtained from the slope in the low-frequency background in dielectric absorption, generally decreases with decreasing temperature due to increasing viscosity (Figure 10). The $\sigma$ values show an Arrhenius dependence in the SmC$_A$* phase, with a lower and higher activation energy above and below 323 K: $E_a$ = 64.0(1) and 72.8(5) kJ/mol for MIX5HFHH6, $E_a$ = 57.6(2) and 70.4(6) kJ/mol for MIX6HFHH6. An increase in activation energy at lower temperatures shows an influence of the glass transition even much above $T_g$. By extrapolating the α-relaxation time to higher temperatures using the MYEGA model, one can estimate the effective activation energy based on a local slope in the activation plot. The effective $E_a$ of the α-relaxation increases on cooling from 37 to 88 kJ/mol between 385 K and 295 K for MIX5HFHH6 and from 16 to 70 kJ/mol between 385 K and 285 K for MIX6HFHH6 (inset in Figure 10). The $E_a$ values determined from ionic conductivity are within the $E_a$ range of the α-relaxation for MIX5HFHH6, but higher for MIX6HFHH6, which indicates that the molecular mobility has only a partial impact on ionic conductivity. This result differs from amorphous polymeric glassformers, where a strong correlation between molecular mobility and ionic conductivity was reported [36].

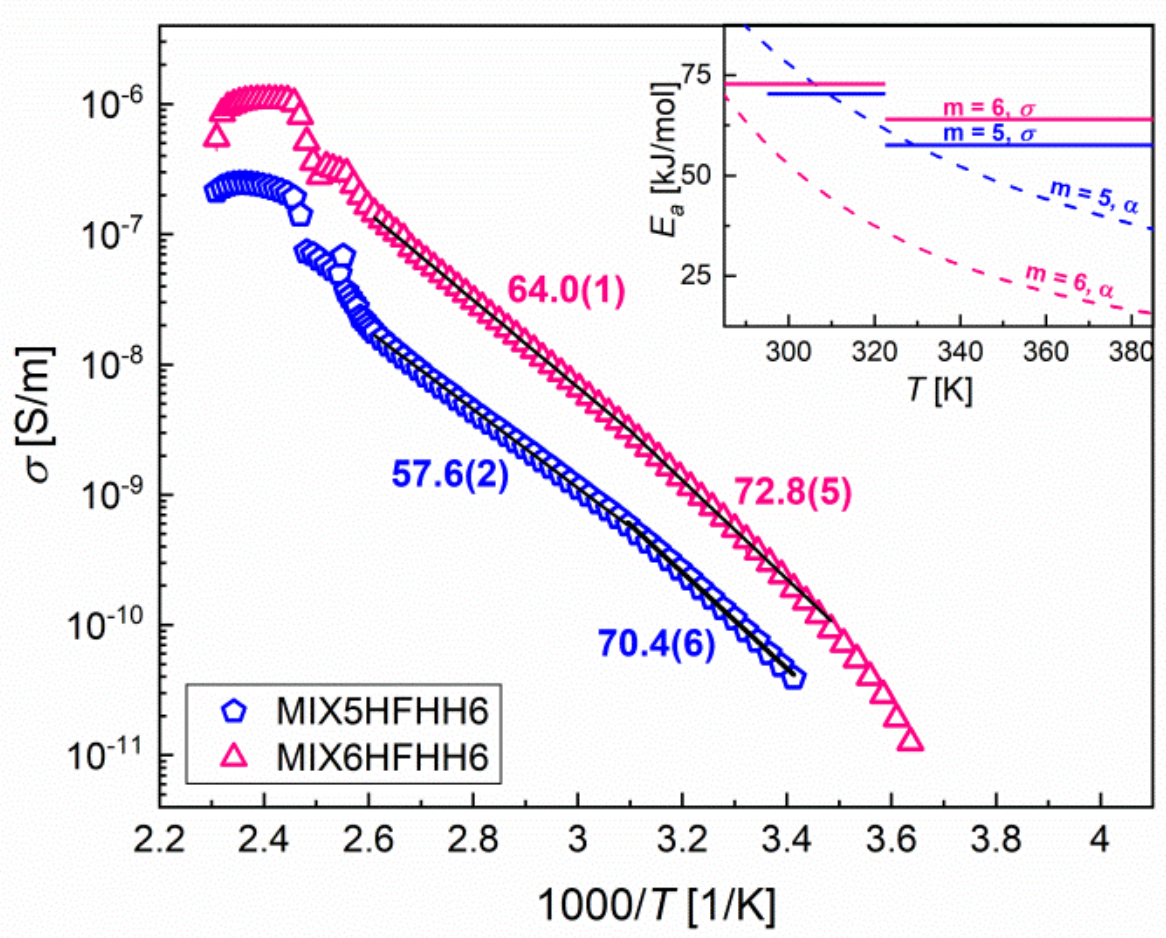

Figure 10. Activation plot of ionic conductivity in MIXmHFHH6. The activation energy is in kJ/mol. The inset shows the effective activation energy calculated by extrapolation of the α-relaxation time to higher temperatures based on the MYEGA model (dashed lines) compared to the activation energy of ionic conductivity (horizontal solid lines).

## 4. Summary and conclusions

The SRL and glassforming properties were reported for two ternary liquid crystalline mixtures, MIXmHFHH6, where m = 5 or 6 denotes the $C_mH_{2m}$ chain length in two components.

- Both mixtures form the SmA*, SmC*, and $SmC_A$* phases; formation of a hexatic phase is hindered by the glass transition. The mixture with m = 5 has better glassforming properties than m = 6, which partially crystallizes on slow cooling.
- The SRL of green light on cooling and red light on heating at 2 K/min is observed in the SmC* phase for both mixtures. This hysteresis is retained for m = 5 at 10 K/min, while the mixture with m = 6 reflects red light both on cooling and heating at a higher rate.
- Thermochromic properties within the $SmC_A$* phase and SRL of mainly blue light in the $SmC_A$* glass are observed for m = 5, while the mixture with m = 6 reflects probably in the infra-red range.
- The glass transition in the $SmC_A$* phase occurs at ca. 235-240 K according to calorimetric results and at ca. 220 K according to the extrapolation of the α-relaxation time to low temperatures by VFT and MYEGA formulas. It is preceded by a maximum in the smectic layer spacing at 248 K.
- A decrease of the dynamic glass transition temperature and fragility index is reported for mixtures compared to their glassforming components. A lower fragility index corresponds to a lower energy difference between constrained and unconstrained states from the MYEGA model in mixtures.

**Acknowledgement:** Aleksandra Deptuch acknowledges the National Science Centre, Poland (grant MINIATURA 7 no. 2023/07/X/ST3/00182) for financial support. The study was carried out using the research infrastructure funded by the European Union in the framework of the Smart Growth Operational Program, Measure 4.2; Grant No. POIR.04.02.00-00-D001/20, "ATOMIN 2.0 – Center for materials research on ATOMic scale for the INnovative economy".

**Conflicts of interest statement:** There are no conflicts to declare.

**Authors' contributions:**
A. Deptuch – conceptualization, investigation, formal analysis, funding acquisition, writing – original draft
Z. Zając – investigation, formal analysis, writing – review and editing
M. Piwowarczyk – investigation, writing – review and editing
A. Drzewicz – investigation, writing – review and editing
M. Kozieł – investigation, writing – review and editing
M. Urbańska – resources, writing – review and editing
E. Juszyńska-Gałązka – investigation, writing – review and editing

**Bibliography**

[1] J.P.F. Lagerwall, F. Giesselmann, Current Topics in Smectic Liquid Crystal Research, ChemPhysChem 7 (2006) 20-45, https://doi.org/10.1002/cphc.200500472.
[2] I. Sage, Thermochromic liquid crystals, Liq. Cryst. 38 (2011) 1551-1561, https://doi.org/10.1080/02678292.2011.631302.
[3] J. Li, H. Takezoe, A. Fukuda, Novel Temperature Dependences of Helical Pitch in Ferroelectric and Antiferroelectric Chiral Smectic Liquid Crystals, Jpn. J. Appl. Phys. 30 (1991) 532-536, https://doi.org/10.1143/JJAP.30.532.
[4] M. Żurowska, R. Dąbrowski, J. Dziaduszek, K. Garbat, M. Filipowicz, M. Tykarska, W. Rejmer, K. Czupryński, A. Spadło, N. Bennis, J.M. Otón, Influence of alkoxy chain length and fluorosubstitution on mesogenic and spectral properties of high tilted antiferroelectric esters, J. Mater. Chem. 21 (2011) 2144-2153, https://doi.org/10.1039/C0JM02015J.
[5] H.S.A. Golicha, M.H. Omar, N.M. Mbithi, Optical Techniques in the Determination of Pitch Lengths in the Cholesteric and Chiral Smectic C Phases, J. Mater. Sci. Eng. 10 (2021) 1-5.
[6] M. Mrukiewicz, M. Cigl, P. Perkowski, J. Karcz, V. Hamplová, A. Bubnov, Dual tunability of selective reflection by light and electric field for self-organizing materials, J. Mol. Liq. 400 (2024) 124540, https://doi.org/10.1016/j.molliq.2024.124540.
[7] M. Krueger, F. Giesselmann, Laser-light diffraction studies on the electric-field response of the helical director configuration in smectic-C* liquid crystals, J. Appl. Phys. 101 (2007) 094102, https://doi.org/10.1063/1.2717609.
[8] M. Mitov, A. Boudet, P. Sopéna, From selective to wide-band light reflection: a simple thermal diffusion in a glassy cholesteric liquid crystal, Eur. Phys. J. B 8 (1999) 327-330, https://doi.org/10.1007/s100510050696.
[9] A. Gujral, L. Yu, M.D. Ediger, Anisotropic organic glasses, Curr. Opin. Solid State Mater. Sci. 22 (2018) 49-57, https://doi.org/10.1016/j.cossms.2017.11.001.
[10] A. Drzewicz, Insight into phase situation and kinetics of cold- and melt crystallization processes of chiral smectogenic liquid crystals, Institute of Nuclear Physics Polish Academy of Sciences, Kraków 2023, https://doi.org/10.48733/978-83-63542-37-5.
[11] A. Deptuch, M. Jasiurkowska-Delaporte, W. Zając, E. Juszyńska-Gałązka, A. Drzewicz, M. Urbańska, Investigation of crystallization kinetics and its relationship with molecular dynamics for chiral fluorinated glassforming smectogen 3F5HPhH6, Phys. Chem. Chem. Phys. 23 (2021) 19795-19810, https://doi.org/10.1039/D1CP02297K.
[12] A. Deptuch, M. Jasiurkowska-Delaporte, M. Urbańska, S. Baran, Kinetics of cold crystallization in two liquid crystalline fluorinated homologues exhibiting the vitrified smectic $C_A$* phase, J. Mol. Liq. 368 (2022) 120612, https://doi.org/10.1016/j.molliq.2022.120612.

[13] W.-L. Tsai, K.-Y. Huang, C.-Y. Hsueh, K.-T. Wang, C.-C. Wen, H.-M. Lai, P.-S. Weng, Synthesis and Liquid Crystal Properties of Chiral Compounds Containing the Core Structure of 6-Hydroxynicotinic Acid or 4-Hydroxyphenylacetic Acid, 53 (2006) 1385-1390, https://doi.org/10.1002/jccs.200600183.
[14] N. Osiecka, Z. Galewski, M. Massalska-Arodź, TOApy program for the thermooptical analysis of phase transitions, Thermochim. Acta 655 (2017) 106-111, https://doi.org/10.1016/j.tca.2017.06.012.
[15] C.A. Schneider, W.S. Rasband, K.W. Eliceiri, NIH Image to ImageJ: 25 years of image analysis, Nat. Methods 9 (2012) 671-675, https://doi.org/10.1038/nmeth.2089.
[16] J.A. Bearden, X-Ray Wavelengths, Rev. Mod. Phys. 39 (1967) 78-124, https://doi.org/10.1103/RevModPhys.39.78.
[17] C.R. Hubbard, National Bureau of Standards Certificate, Standard Reference Material 675, Low 2θ (Large d-Spacing) Standard for X-Ray Powder Diffraction, https://tsapps.nist.gov/srmext/certificates/675.pdf, access on March 2024.
[18] T. Roisnel, J. Rodríquez-Carvajal, WinPLOTR: A Windows Tool for Powder Diffraction Pattern Analysis, Mater. Sci. Forum 378-381 (2001) 118-123, https://doi.org/10.4028/www.scientific.net/MSF.378-381.118.
[19] E. Ghanbari, S.J. Picken, J.H. van Esch, Analysis of differential scanning calorimetry (DSC): determining the transition temperatures, and enthalpy and heat capacity changes in multicomponent systems by analytical model fitting, J. Therm. Anal. Calorim. 148 (2023) 12393-12409, https://doi.org/10.1007/s10973-023-12356-1.
[20] TA Instruments, Overview of Glass Transition Analysis by Differential Scanning Calorimetry, https://www.tainstruments.com/pdf/literature/TA443.pdf, access on 27th August 2025.
[21] J.M. Seddon, Structural studies of Liquid Crystals by X-ray Diffraction, in D. Demus, J. Goodby, G.W. Gray, H.-W. Spiess, V. Vill (Eds.), Handbook of Liquid Crystals, WILEY-VCH Verlag GmbH, Weinheim 1998.
[22] P. Davidson, Selected Topics in X-Ray Scattering by Liquid-Crystalline Polymers, in D.M.P. Mingos (Ed.), Liquid Crystals II. Structure and Bonding, vol 95, Springer, Berlin Heidelberg 1999, https://doi.org/10.1007/3-540-68118-3_1.
[23] W. Massa, Crystal Structure Determination, Springer-Verlag, Berlin Heidelberg 2000, https://doi.org/10.1007/978-3-662-04248-9.
[24] Y. Takanishi, K. Miyachi, S. Yoshida, B. Jin, H. Yin, K. Ishikawa, H. Takezoe, A. Fukuda, Stability of antiferroelectricity and molecular reorientation in the hexatic smectic $I_A$* phase as studied by X-ray diffraction and NMR spectroscopy, J. Mater. Chem. 8 (1998) 1133-1138, https://doi.org/10.1039/A707920F.
[25] A. Bubnov, V. Novotná, V. Hamplová, M. Kašpar, M. Glogarová, J. Mol. Struc. 892 (2008) 151-157, https://doi.org/10.1016/j.molstruc.2008.05.016.
[26] S. Havriliak, S. Negami, A complex plane analysis of α-dispersions in some polymer systems, J. Polym. Sci. C: Polymer Symposia 14 (1966) 99-117, https://doi.org/10.1002/polc.5070140111.
[27] W. Haase, S. Wróbel (Eds.), Relaxation phenomena. Liquid crystals, magnetic systems, polymers, high-Tc superconductors, metallic glasses, Springer-Verlag, Berlin Heidelberg 2003, https://doi.org/10.1007/978-3-662-09747-2.
[28] F. Kremer, A. Schönhals (Eds.), Broadband Dielectric Spectroscopy, Springer-Verlag, Berlin Heidelberg 2003, https://doi.org/10.1007/978-3-642-56120-7.
[29] M. Buivydas, F. Gouda, G. Andersson, S.T. Lagerwall, B. Stebler, J. Bömelburg, G. Heppke, B. Gestblom, Collective and non-collective excitations in antiferroelectric and ferrielectric liquid crystals studied by dielectric relaxation spectroscopy and electro-optic measurements, Liq. Cryst. 23 (1997) 723-739, https://doi.org/10.1080/026782997208000.
[30] R. Böhmer, K.L. Ngai, C.A. Angell, D.J. Plazek, Nonexponential relaxations in strong and fragile glass formers, J. Chem. Phys. 99 (1993) 4201-4209, https://doi.org/10.1063/1.466117.
[31] D. Georgopoulos, S. Kripotou, E. Argyraki, A. Kyritsis, P. Pissis, Study of Isothermal Crystallization Kinetics of 5CB with Differential Scanning Calorimetry and Broadband Dielectric Spectroscopy, Mol. Cryst. Liq. Cryst. 611 (2015) 197-207, https://doi.org/10.1080/15421406.2015.1030259.
[32] A. Drozd-Rzoska, S.J. Rzoska, S. Starzonek, New scaling paradigm for dynamics in glass-forming systems, Prog. Mater. Sci. 134 (2023) 101074, https://doi.org/10.1016/j.pmatsci.2023.101074.

[33] K.L. Ngai, M. Paluch, Classification of secondary relaxation in glass-formers based on dynamic properties, J. Chem. Phys. 120 (2004) 857-873, https://doi.org/10.1063/1.1630295.
[34] M. Tarnacka, K. Adrjanowicz, E. Kaminska, K. Kaminski, K. Grzybowska, K. Kolodziejczyk, P. Wlodarczyk, L. Hawelek, G. Garbacz, A. Kocot, M. Paluch, Molecular dynamics of itraconazole at ambient and high pressure, Phys. Chem. Chem. Phys. 15 (2013) 20742-20752, https://doi.org/10.1039/C3CP52643G.
[35] J.C. Mauro, Y. Yue, A.J. Ellison, P.K. Gupta, D.C. Allan, Viscosity of glass-forming liquids, Proc. Natl. Acad. Sci. 106 (2009) 19780-19784, https://doi.org/10.1073/pnas.0911705106.
[36] K.N. Raftopoulos, I. Łukaszewska, S. Lalik, P. Zając, A. Bukowczan, E. Hebda, M. Marzec, K. Pielichowski, Structure-Glass Transition Relationships in Non-Isocyanate Polyhydroxyurethanes, Molecules 29 (2024) 4057, https://doi.org/10.3390/molecules29174057.

**Selective reflection of light in glassforming ternary liquid crystalline mixtures**

Aleksandra Deptuch[1,*], Zuzanna Zając[2], Marcin Piwowarczyk[1], Anna Drzewicz[1], Marcin Kozieł[3], Magdalena Urbańska[4], Ewa Juszyńska-Gałązka[1,5]

[1] Institute of Nuclear Physics, Polish Academy of Sciences, Radzikowskiego 152, PL-31342 Kraków, Poland

[2] Faculty of Materials Science and Ceramics, AGH University of Cracow, PL-30059 Kraków, Mickiewicza 30, Poland

[3] Faculty of Chemistry, Jagiellonian University, Gronostajowa 2, PL-30387, Kraków, Poland

[4] Institute of Chemistry, Military University of Technology, Kaliskiego 2, PL-00908 Warsaw, Poland

[5] Research Center for Thermal and Entropic Science, Graduate School of Science, Osaka University, 560-0043 Osaka, Japan

*corresponding author, aleksandra.deptuch@ifj.edu.pl

# Supplementary materials

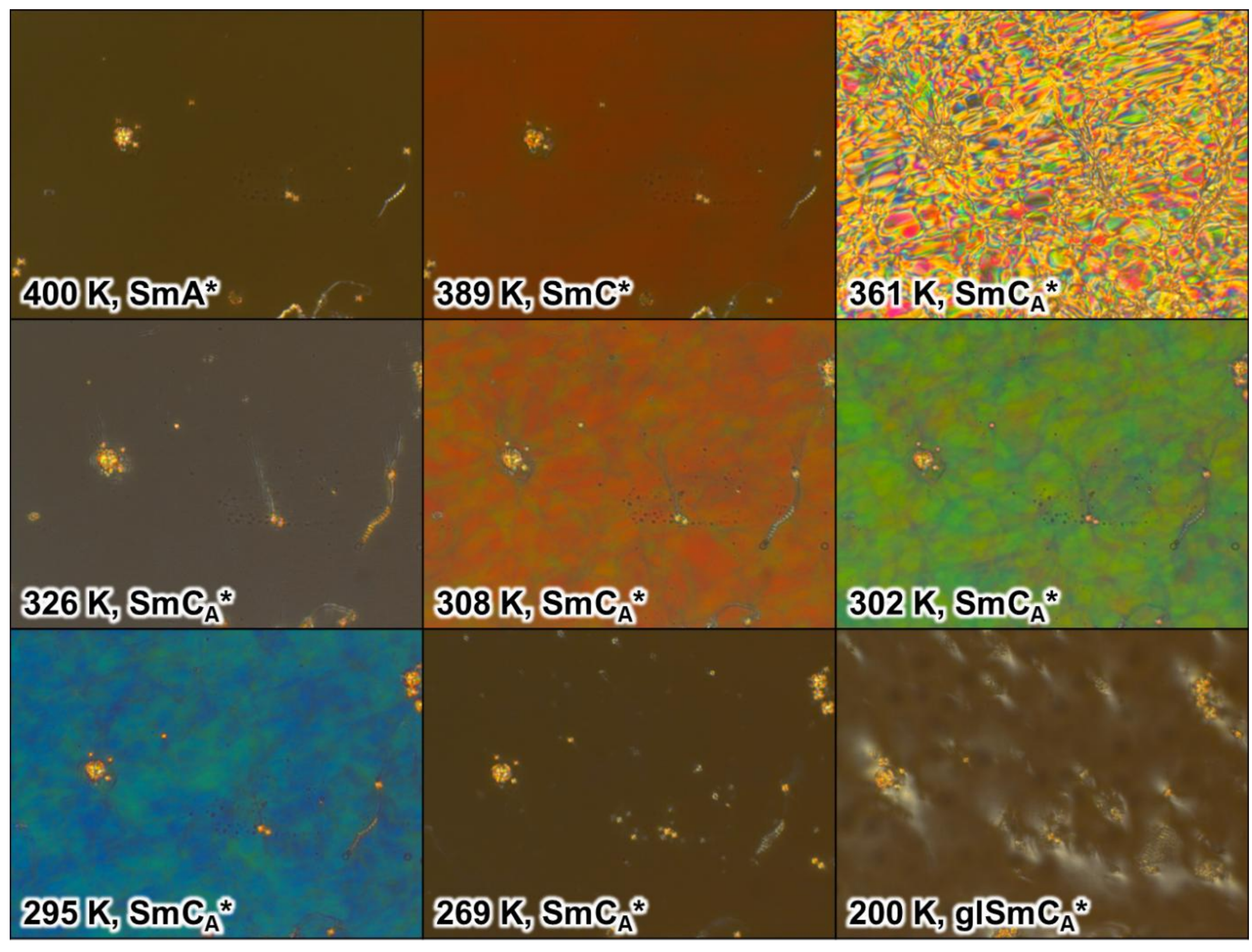

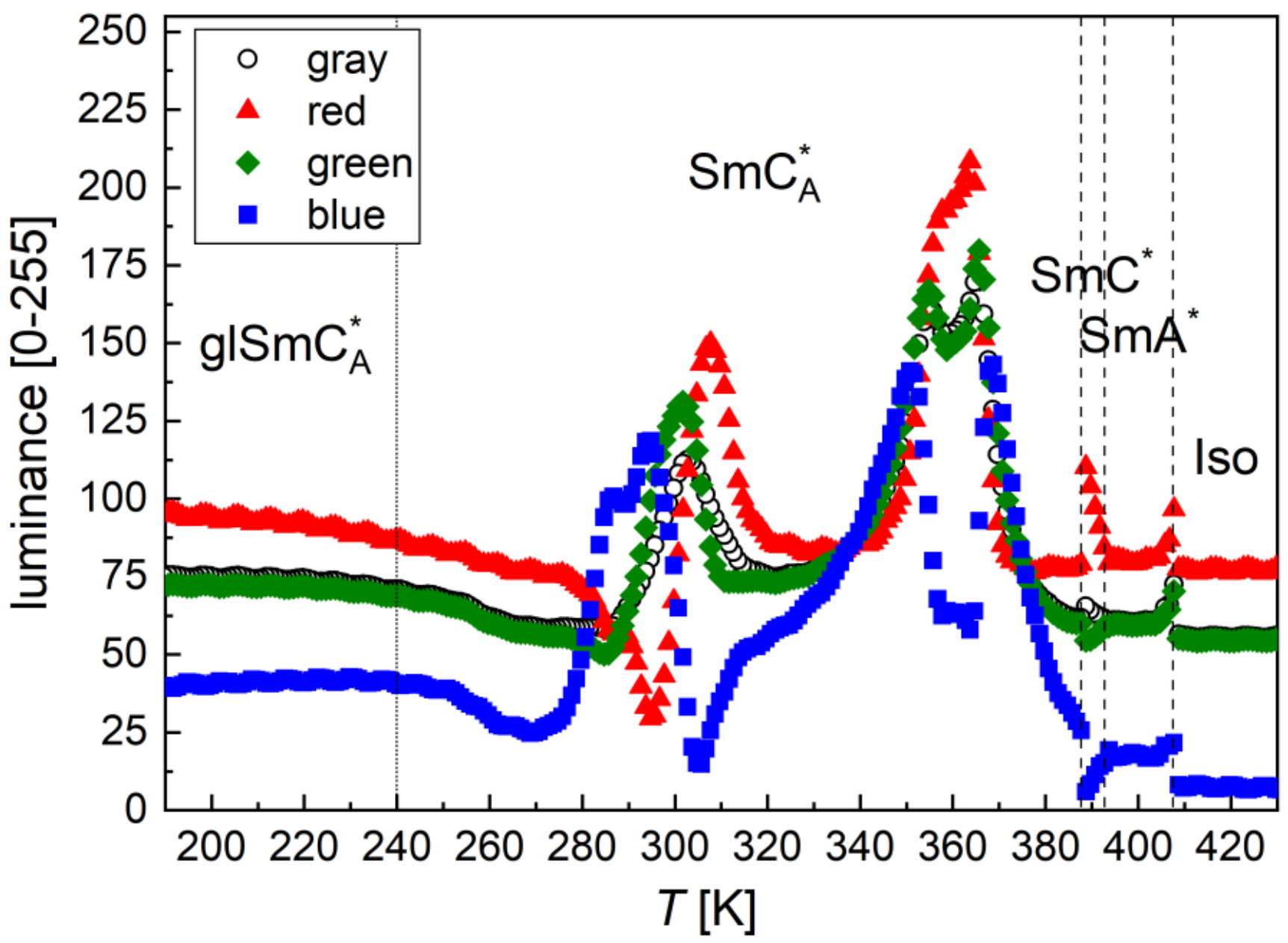

Figure S1. Representative POM textures (622 × 466 μm$^2$) of MIX5HFHH6 collected at the 10 K/min cooling rate in the transmission mode as well as the red, green, blue components and weighted total luminance of each texture. The glass transition temperature is based on DSC results.

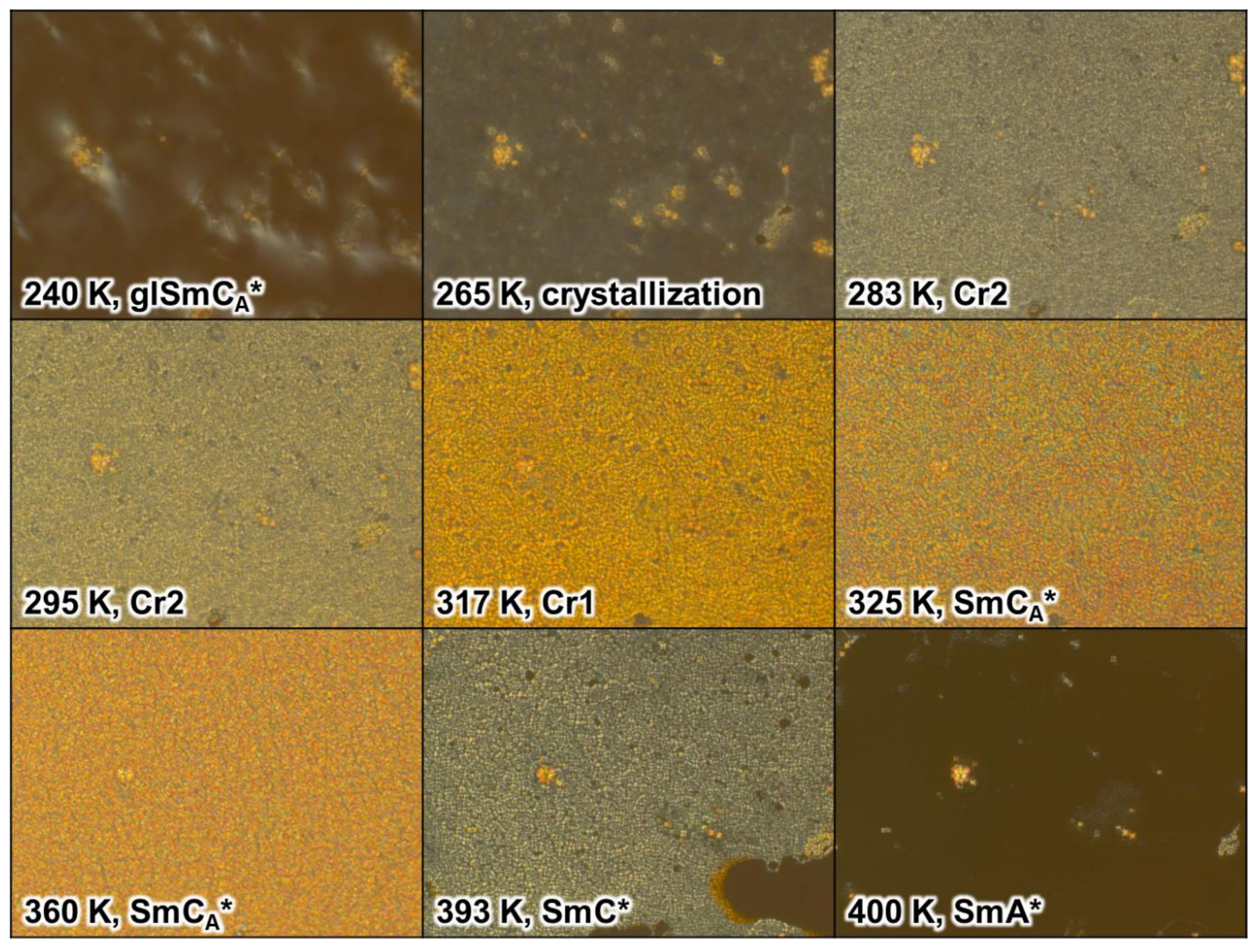

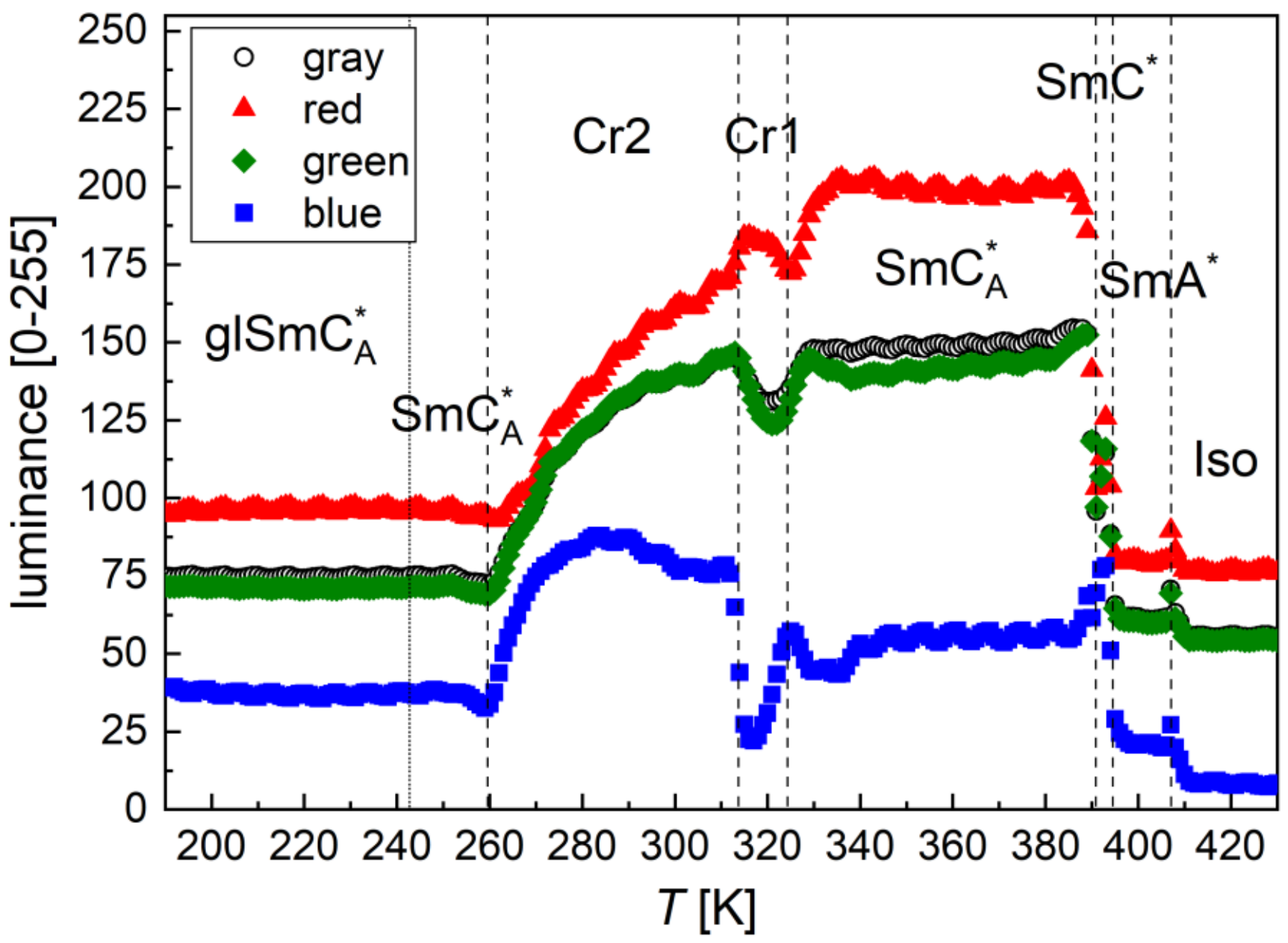

Figure S2. Representative POM textures (622 × 466 μm$^2$) of MIX5HFHH6 collected at the 10 K/min heating rate in the transmission mode as well as the red, green, blue components and weighted total luminance of each texture. The glass transition temperature is based on DSC results.

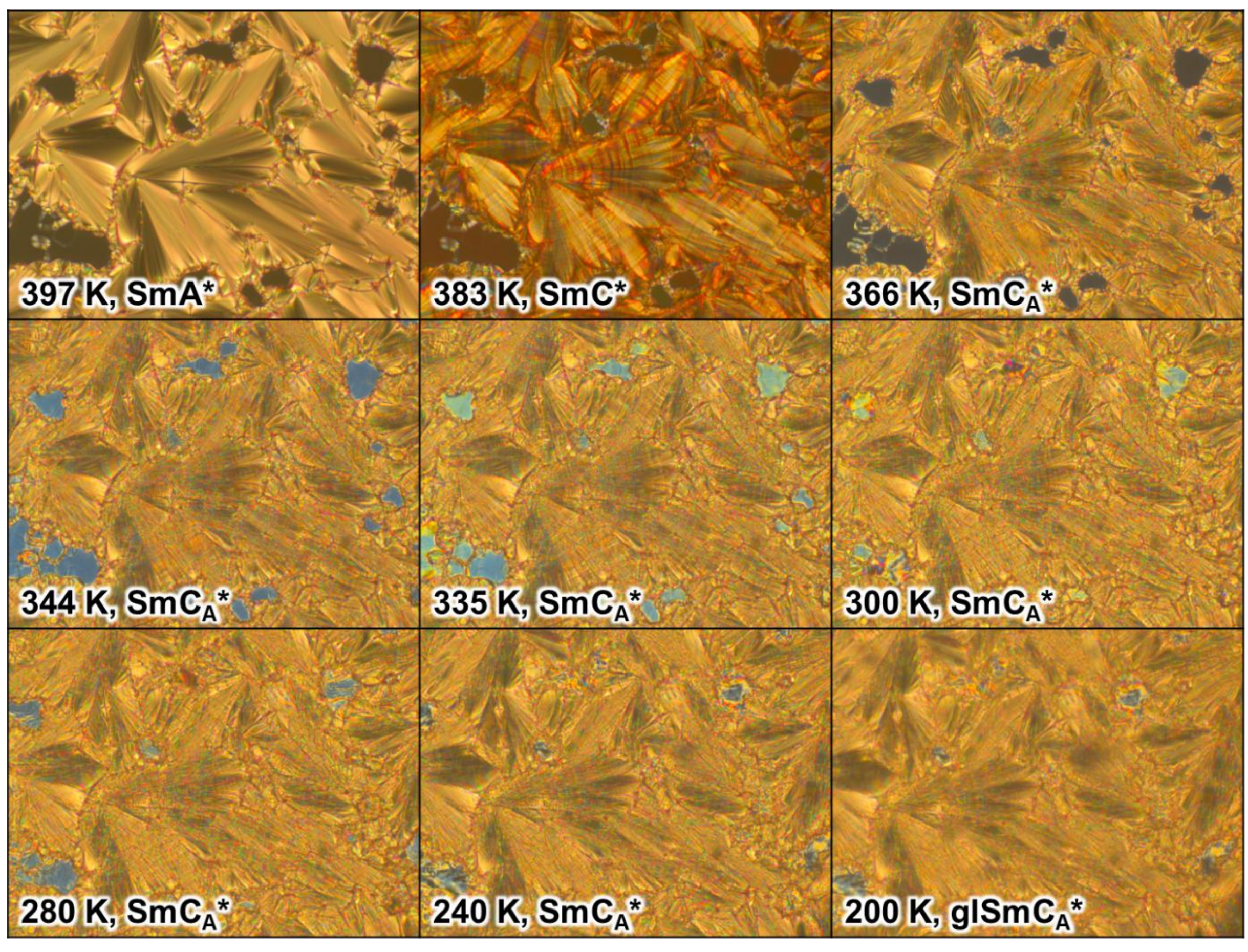

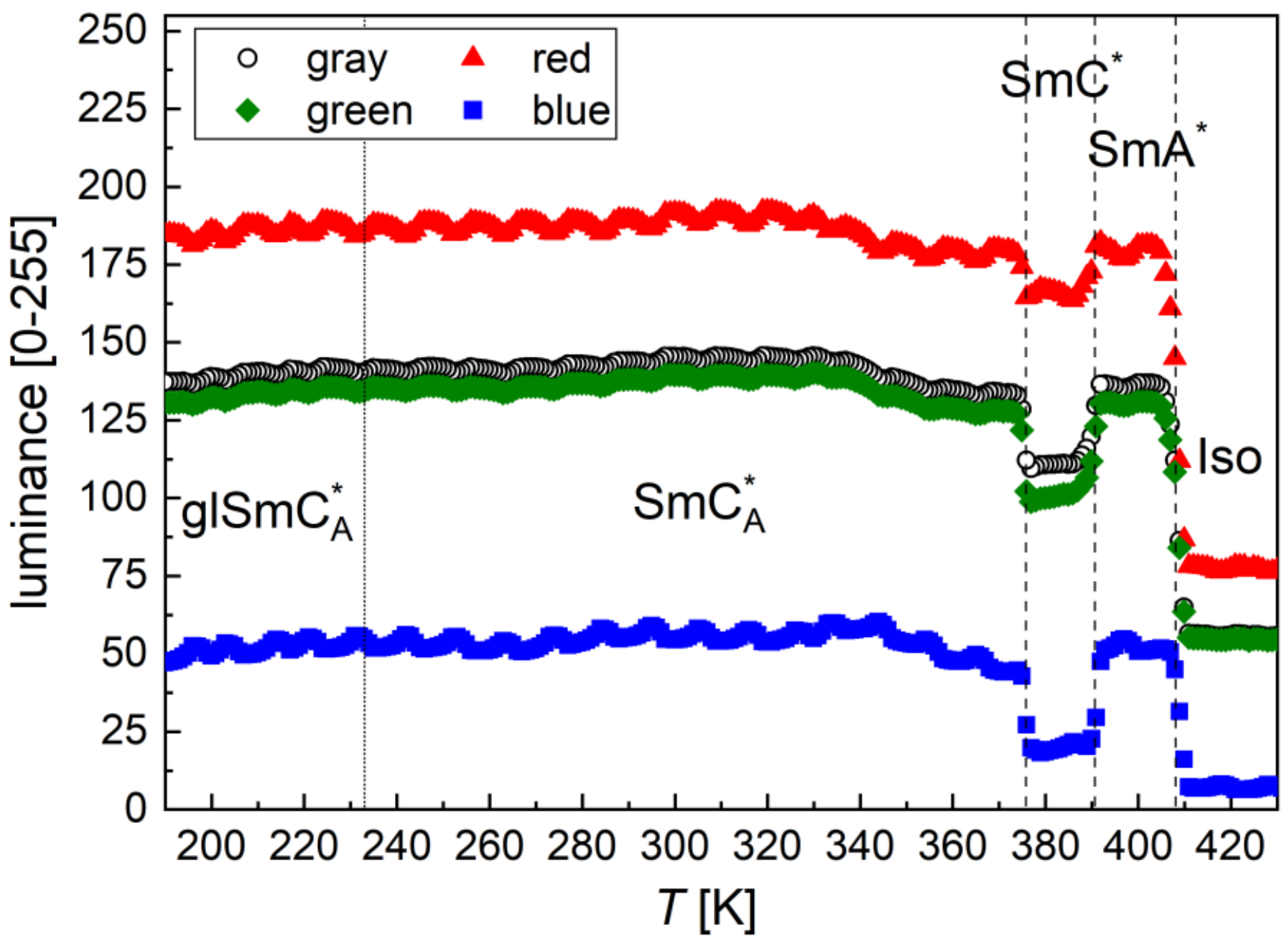

Figure S3. Representative POM textures (622 × 466 μm$^2$) of MIX6HFHH6 collected at the 10 K/min cooling rate in the transmission mode as well as the red, green, blue components and weighted total luminance of each texture. The glass transition temperature is based on DSC results.

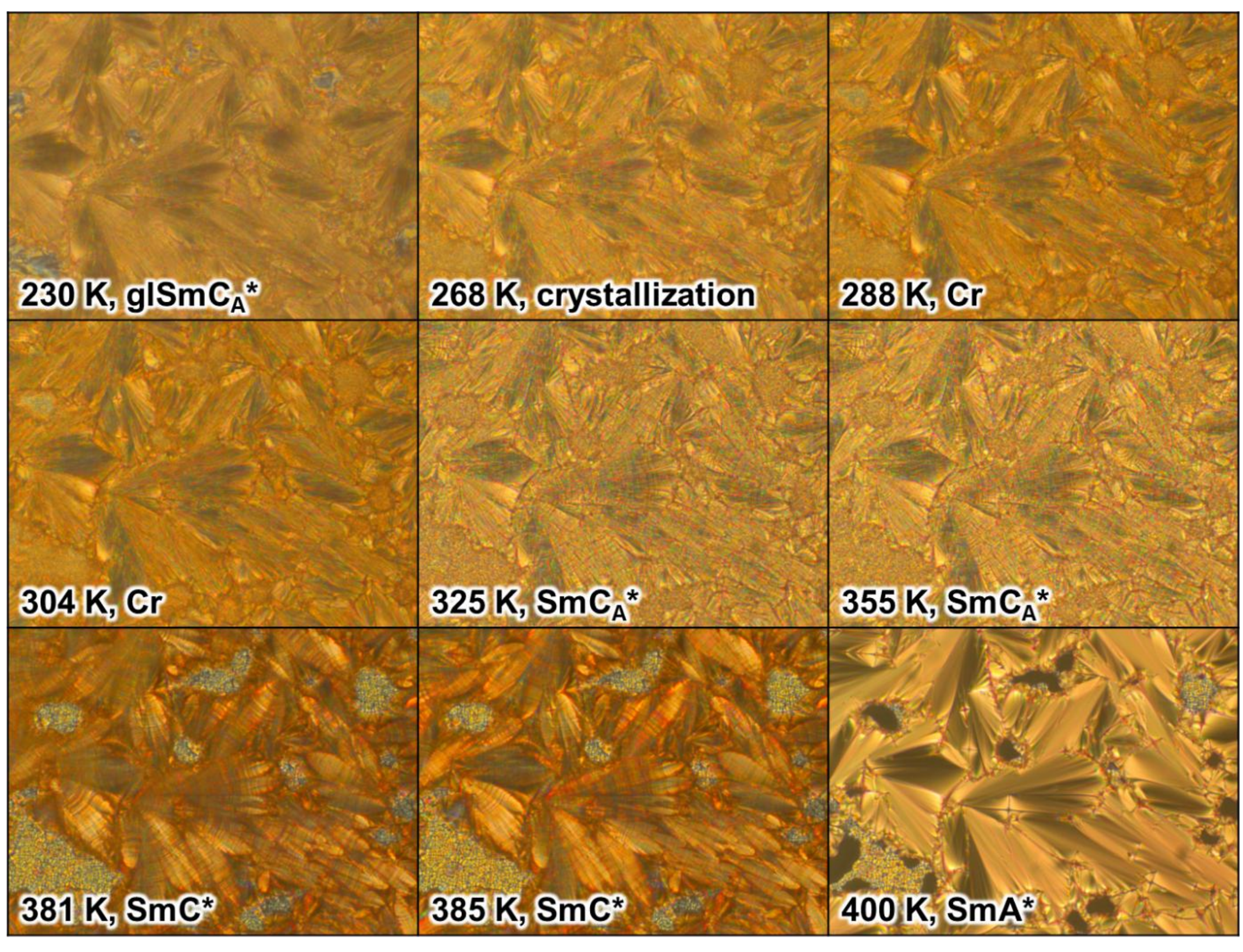

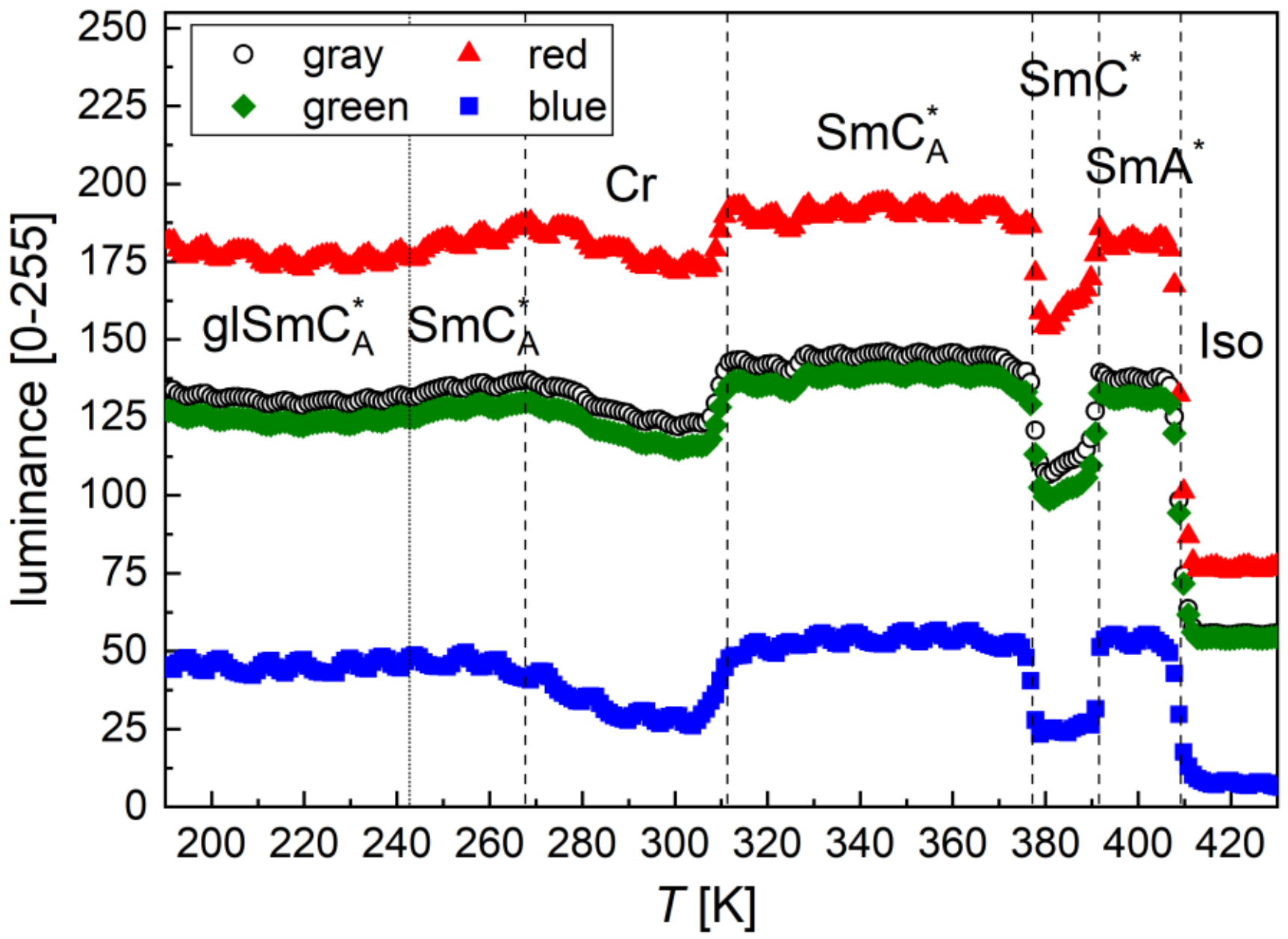

Figure S4. Representative POM textures ($622 \times 466\ \mu m^2$) of MIX6HFHH6 collected at the 10 K/min heating rate in the transmission mode as well as the red, green, blue components and weighted total luminance of each texture. The glass transition temperature is based on DSC results.

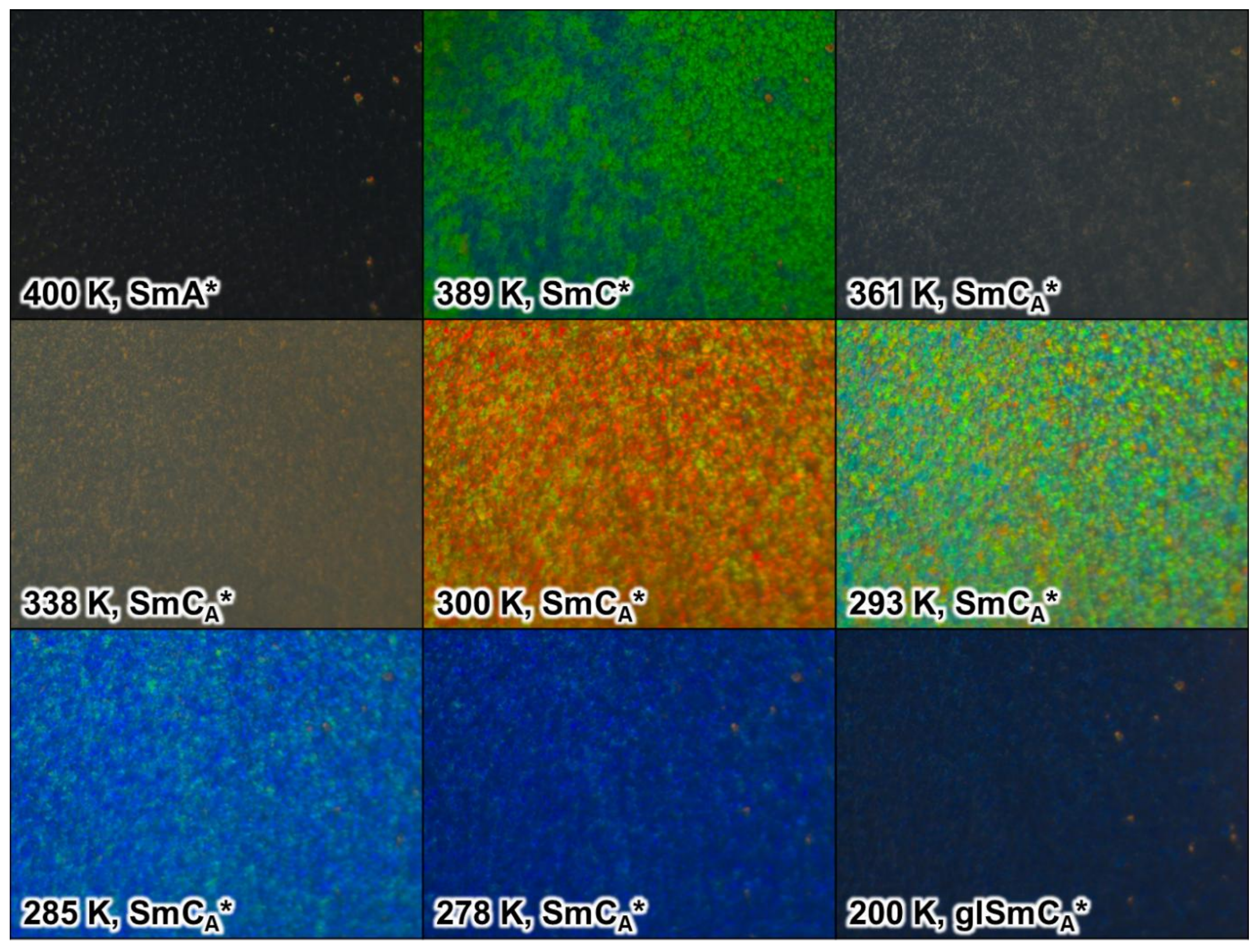

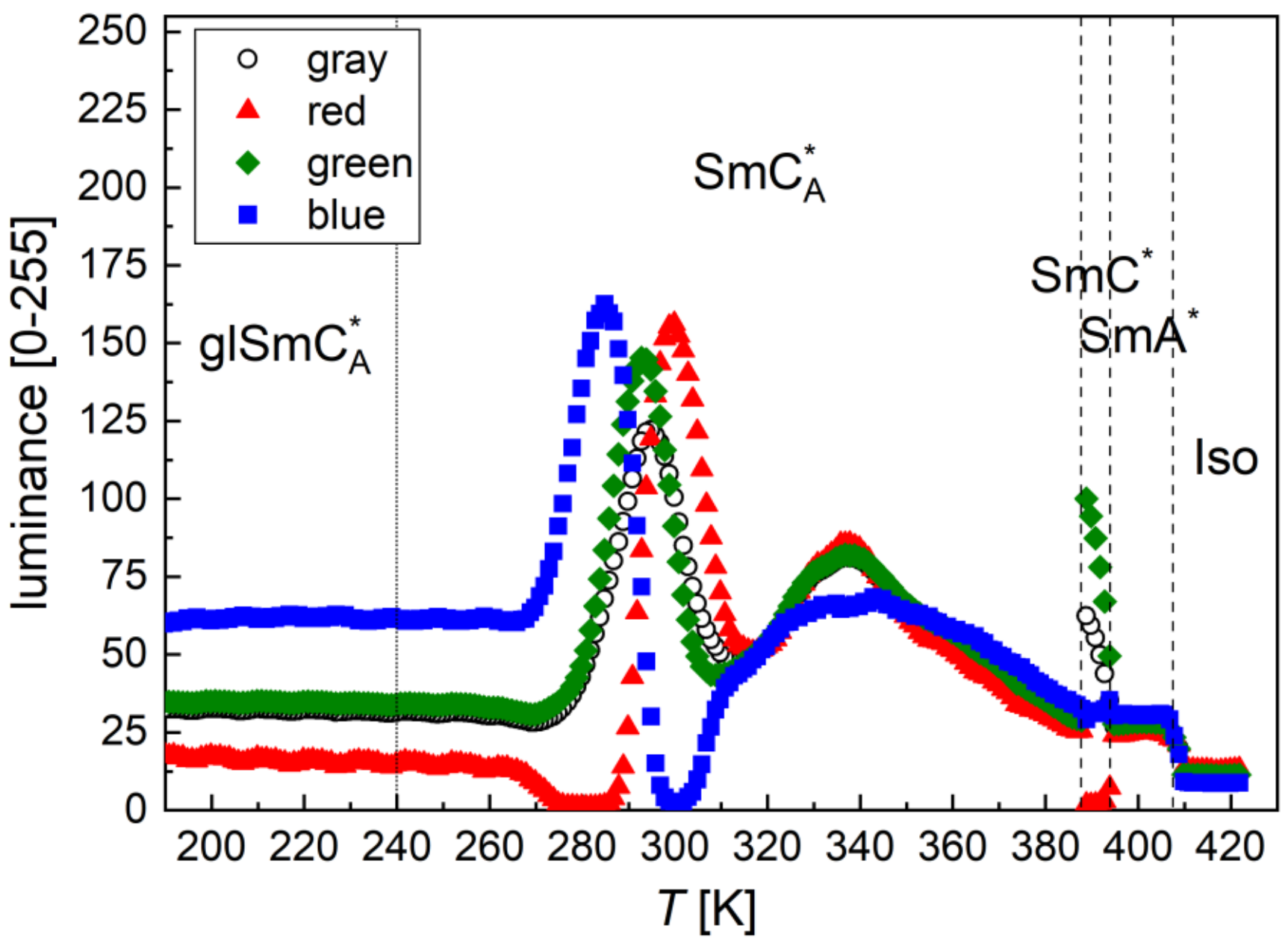

Figure S5. Representative POM textures (622 × 466 μm$^2$) of MIX5HFHH6 collected at the 10 K/min cooling rate in the reflection mode as well as the red, green, blue components and weighted total luminance of each texture. The glass transition temperature is based on DSC results.

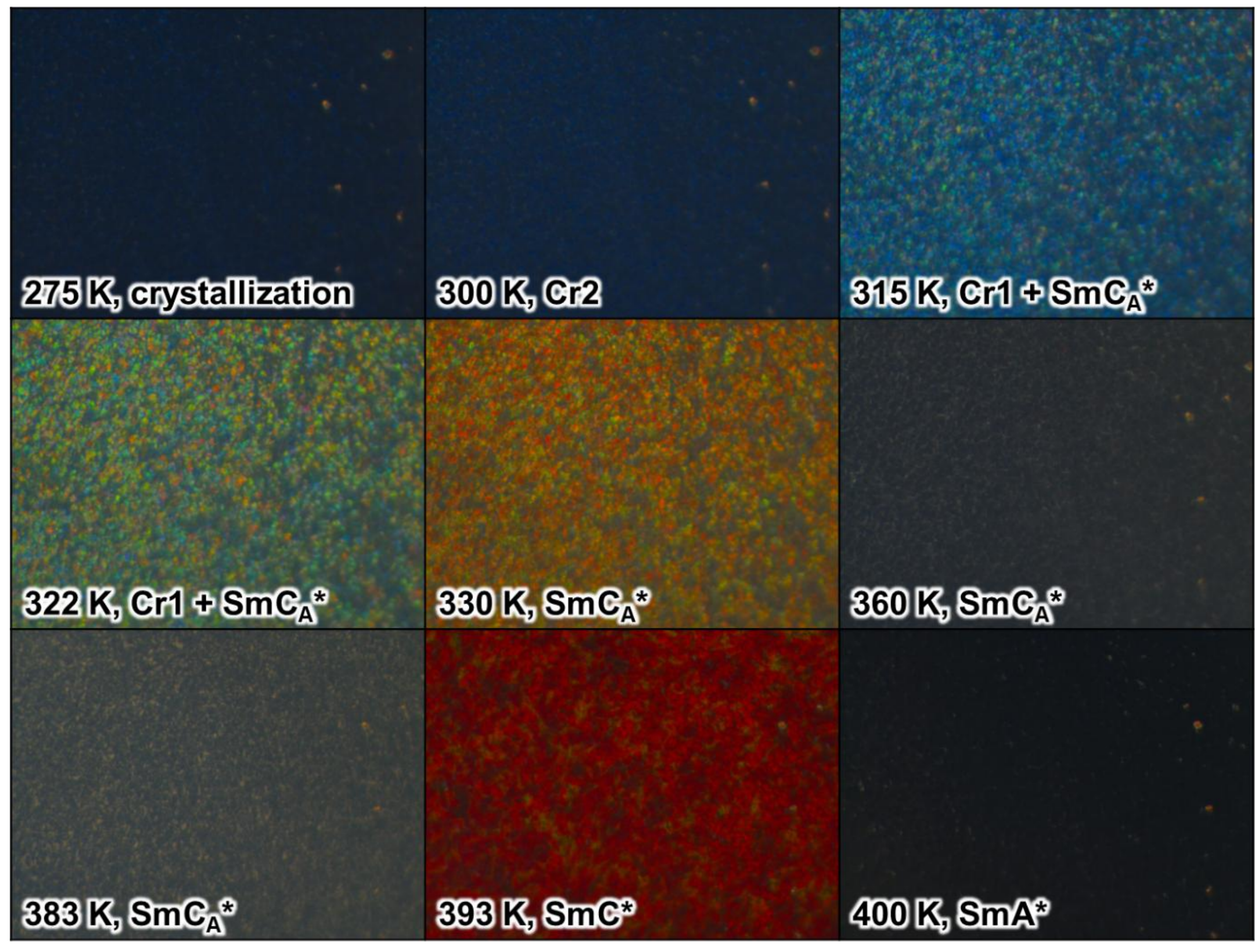

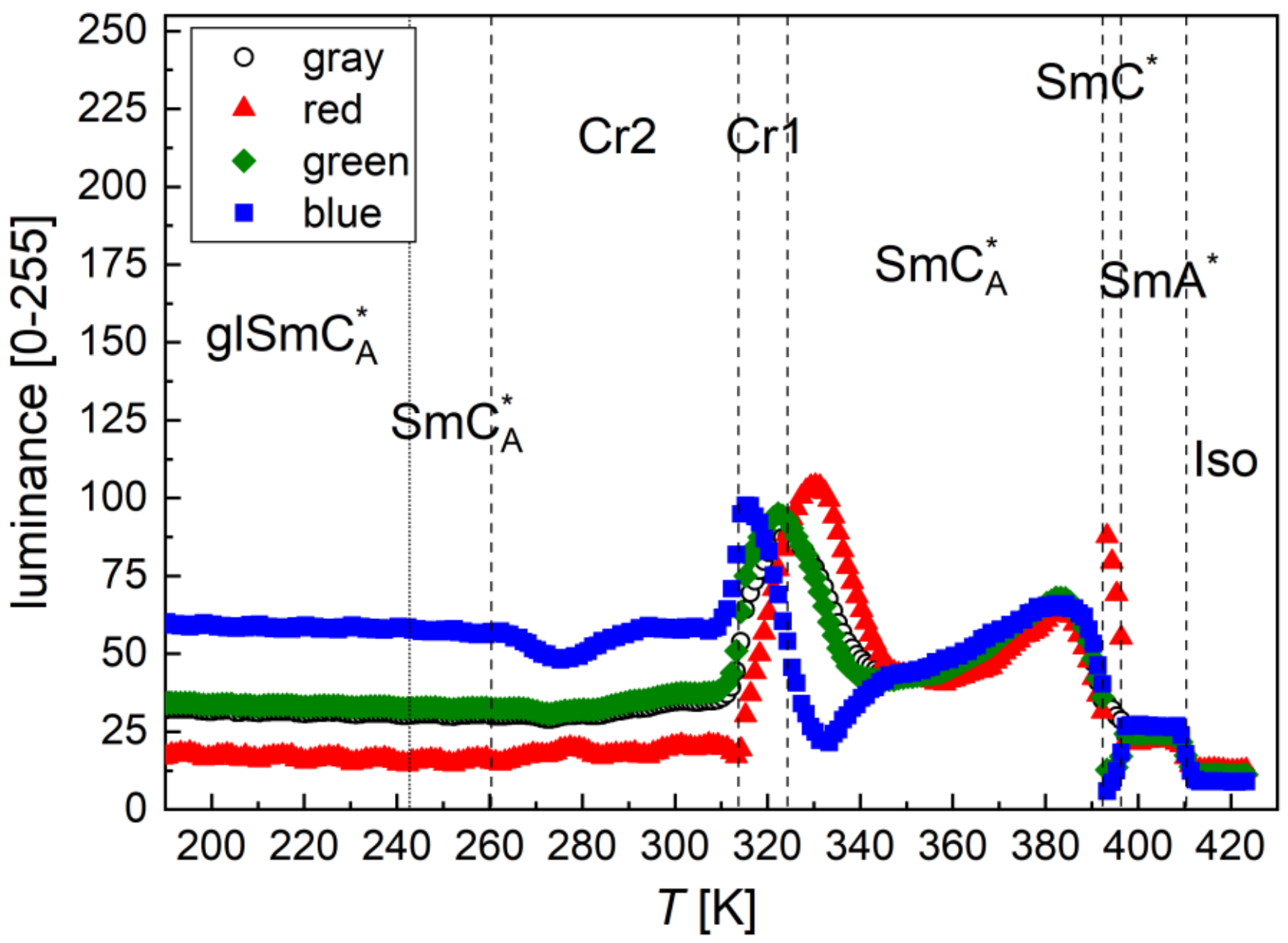

Figure S6. Representative POM textures (622 × 466 μm$^2$) of MIX5HFHH6 collected at the 10 K/min heating rate in the reflection mode as well as the red, green, blue components and weighted total luminance of each texture. The glass transition temperature is based on DSC results.

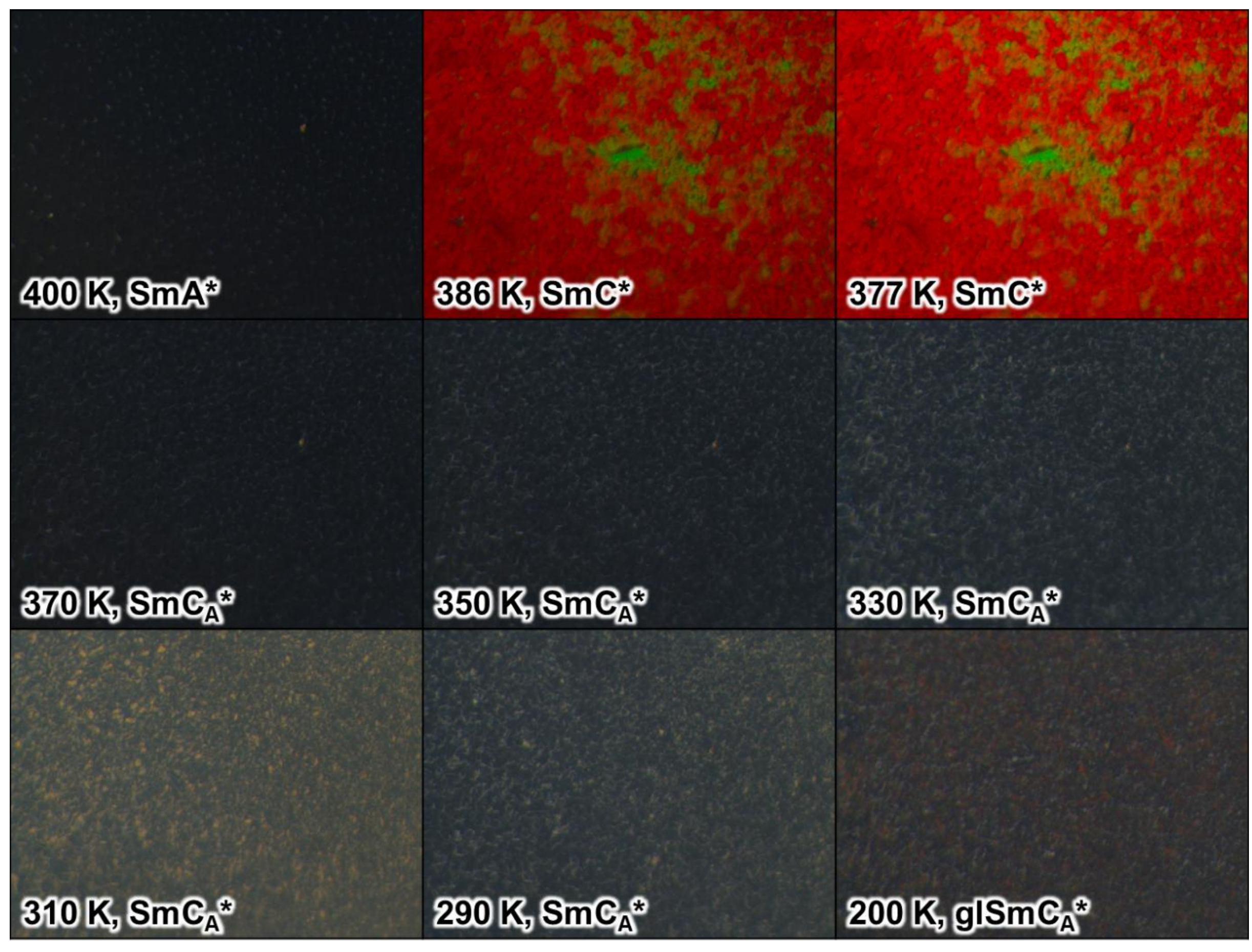

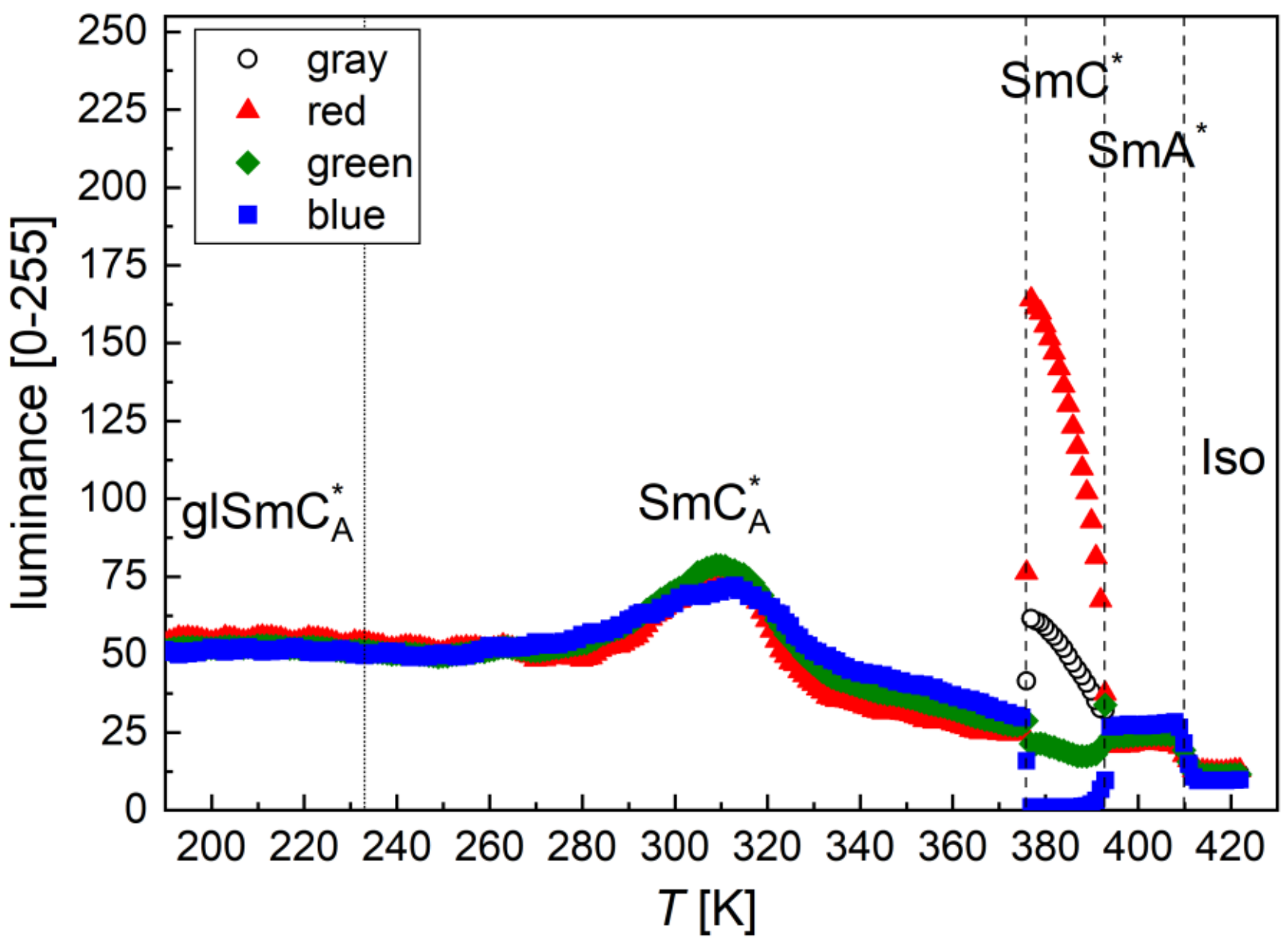

Figure S7. Representative POM textures (622 × 466 μm$^2$) of MIX6HFHH6 collected at the 10 K/min cooling rate in the reflection mode as well as the red, green, blue components and weighted total luminance of each texture. The glass transition temperature is based on DSC results.

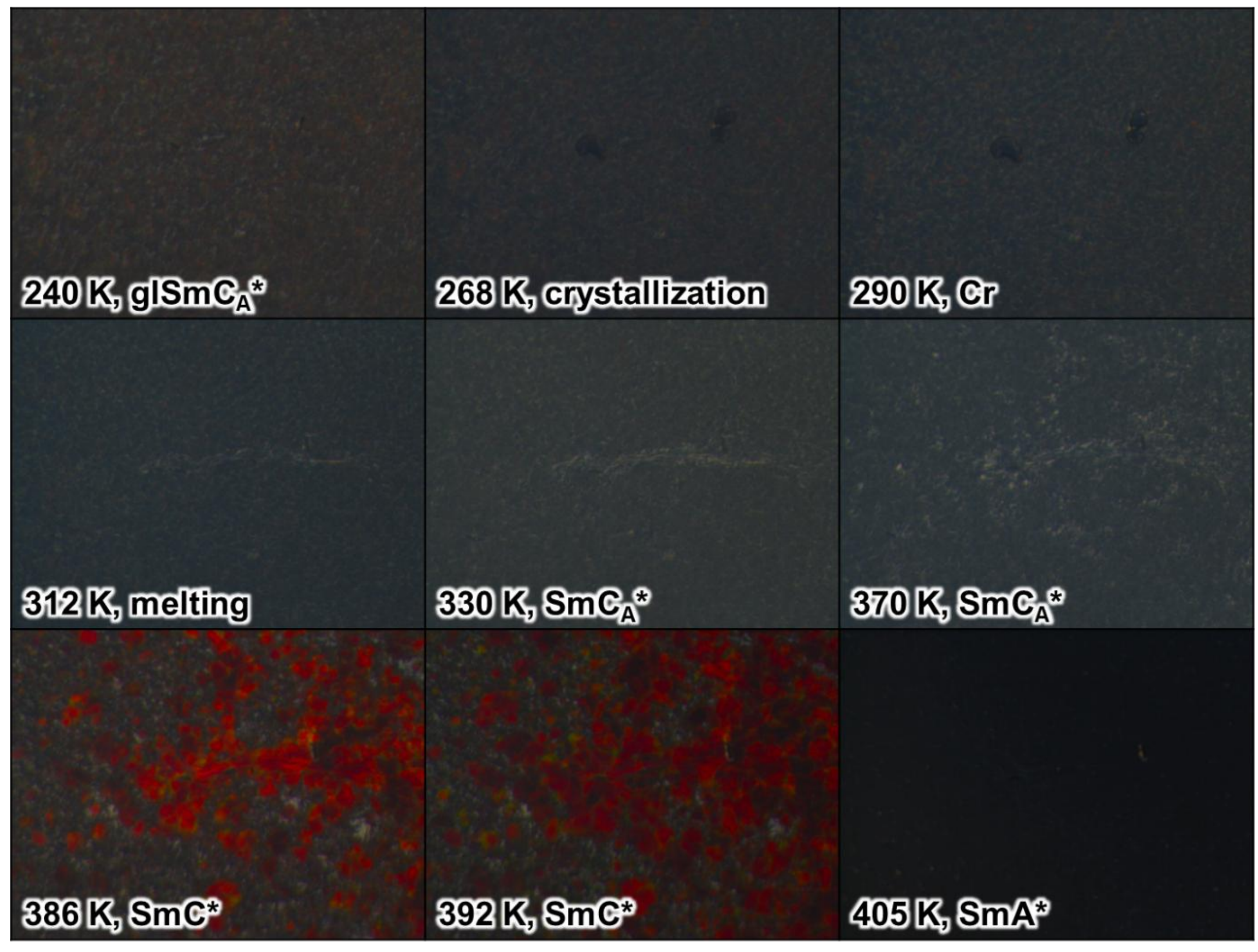

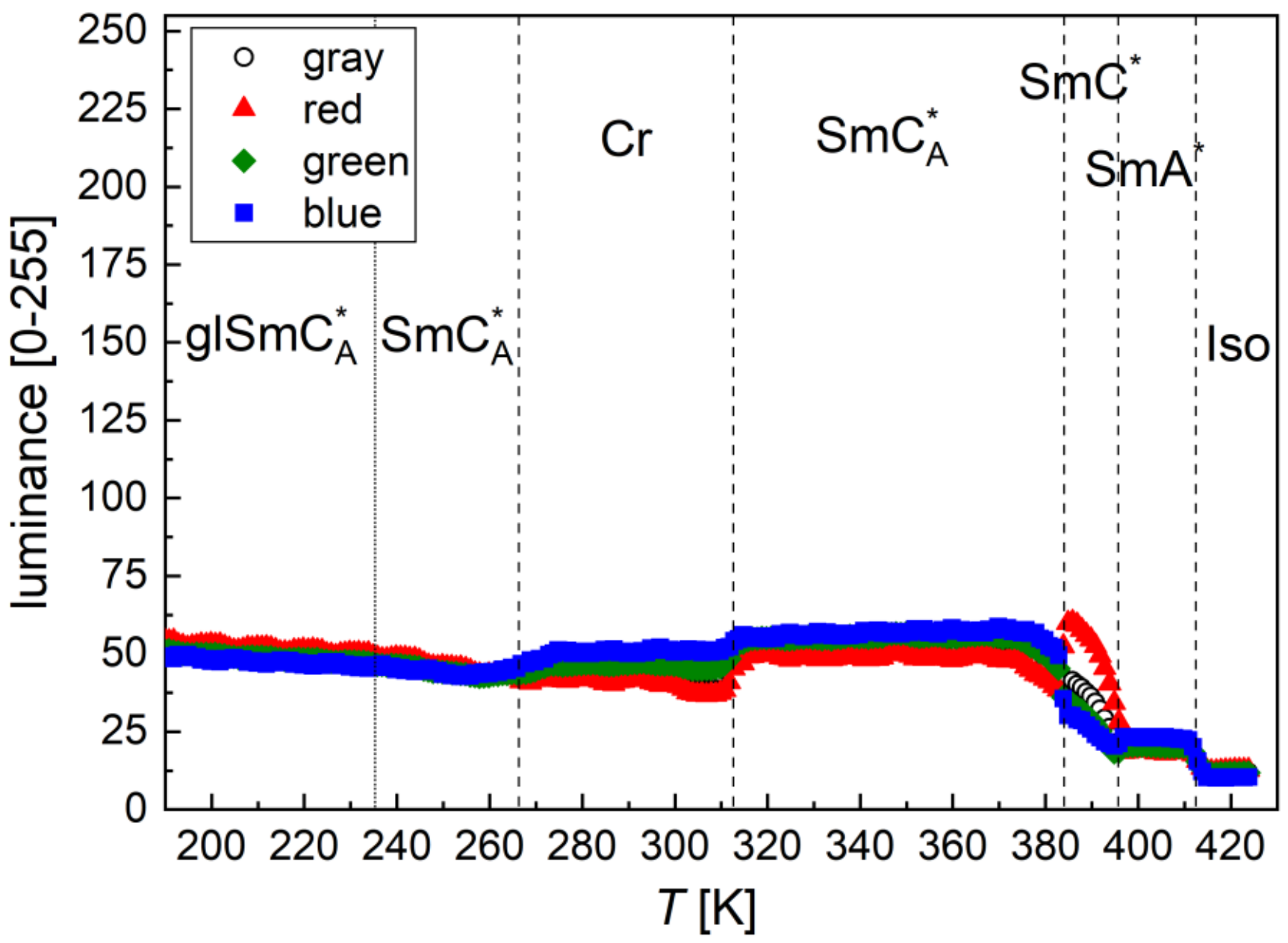

Figure S8. Representative POM textures (622 × 466 μm$^2$) of MIX6HFHH6 collected at the 10 K/min heating rate in the reflection mode as well as the red, green, blue components and weighted total luminance of each texture. The glass transition temperature is based on DSC results.

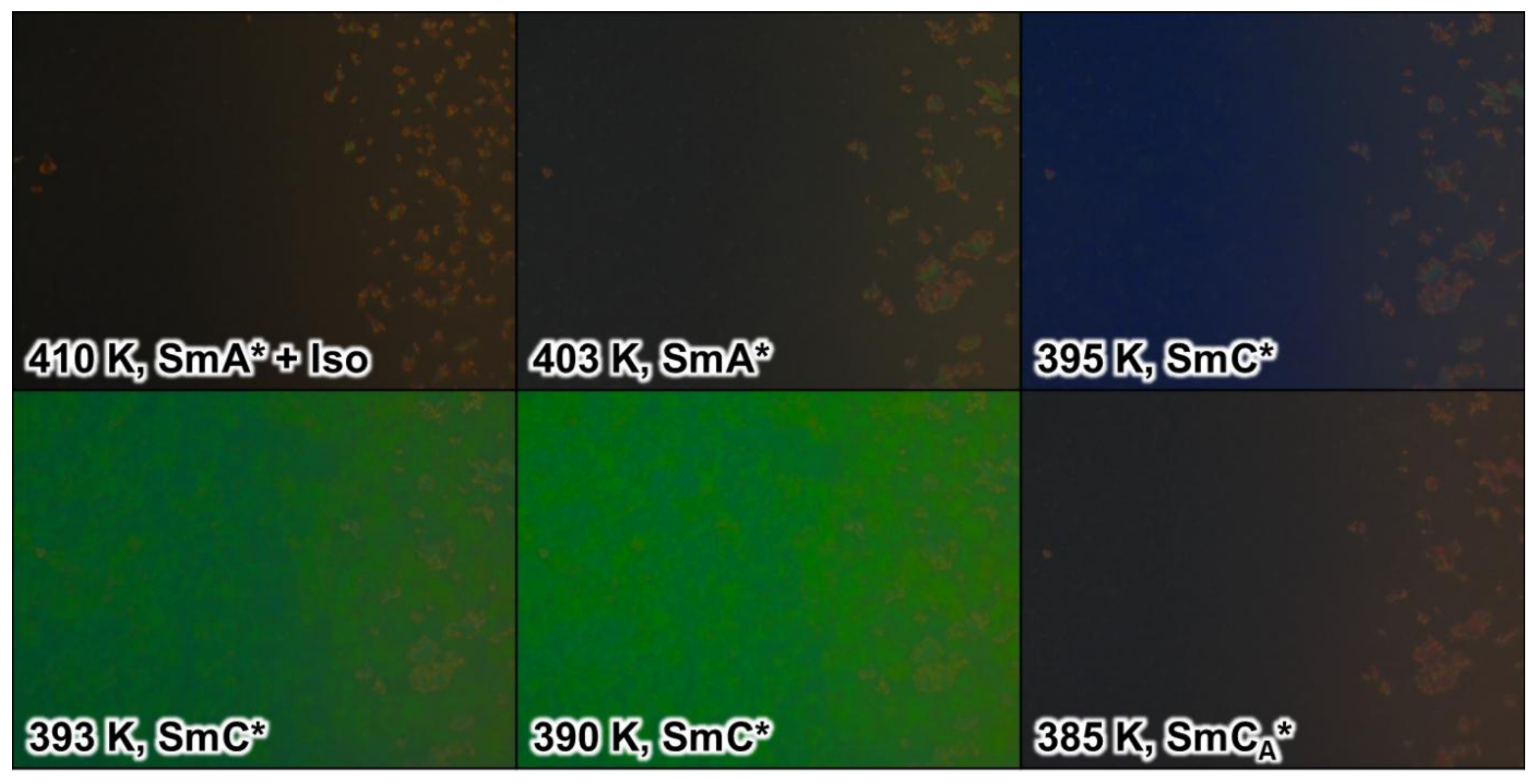

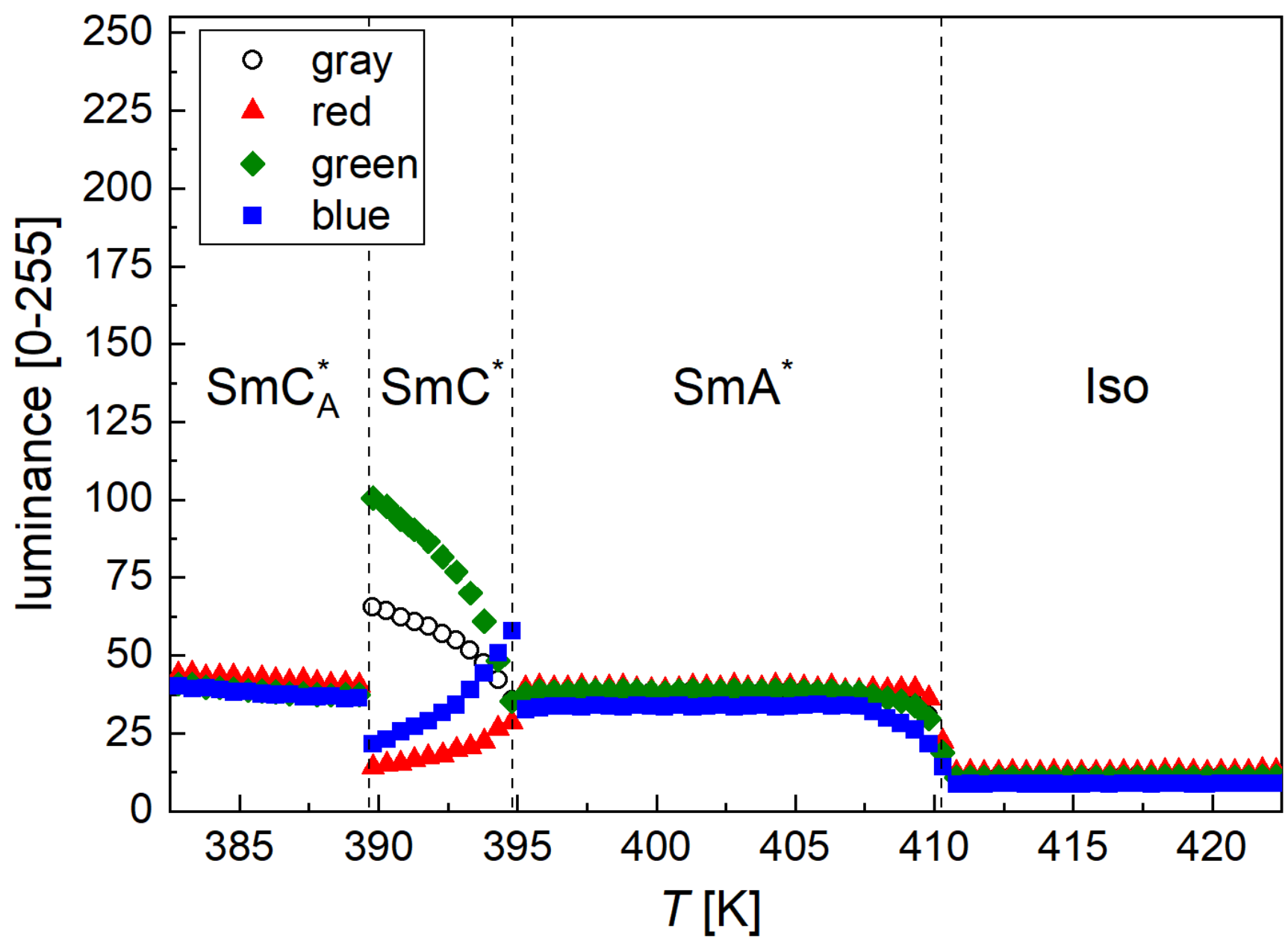

Figure S9. Representative POM textures (622 × 466 μm$^2$) of MIX5HFHH6 collected at the 2 K/min cooling rate in the reflection mode as well as the red, green, blue components and weighted total luminance of each texture.

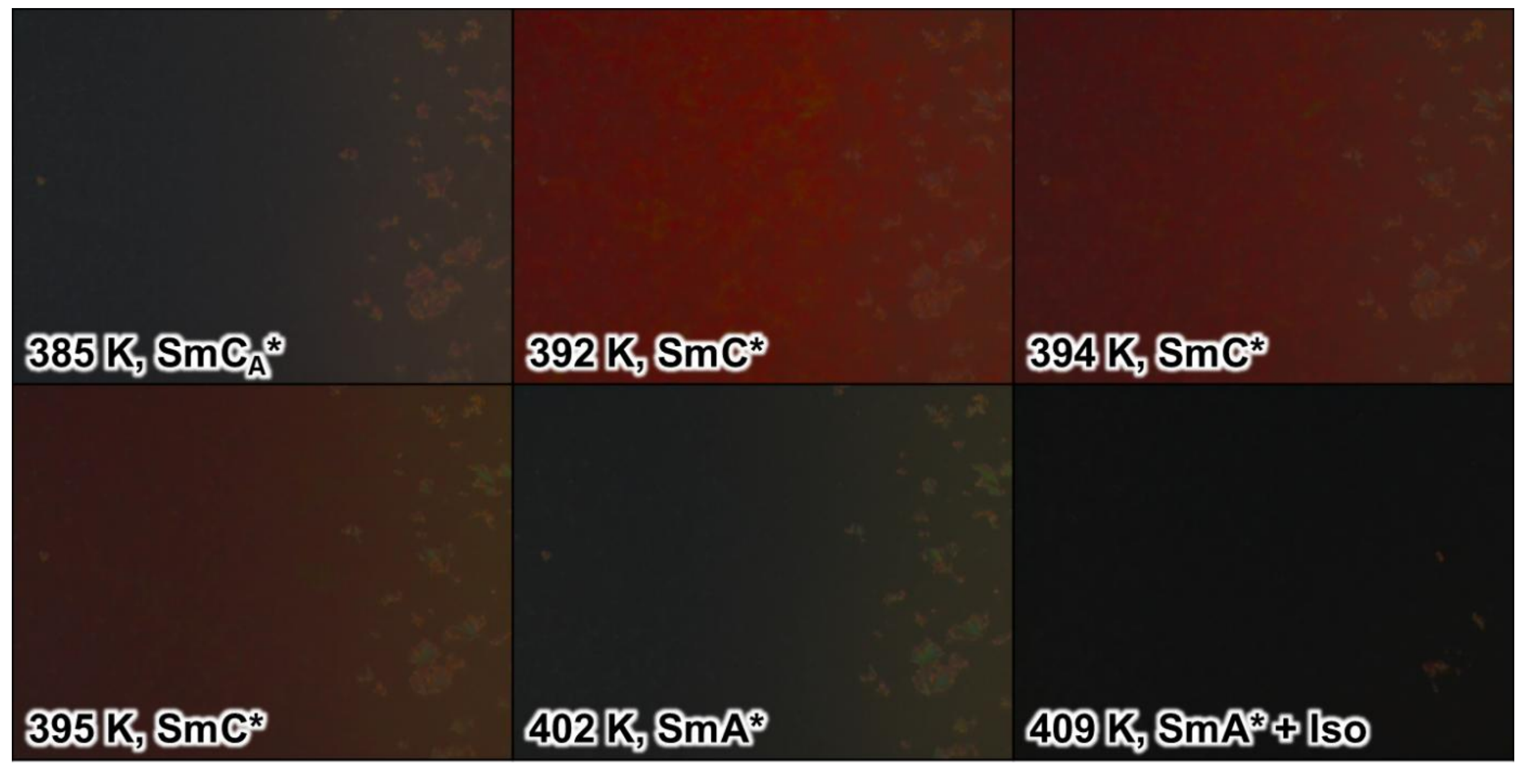

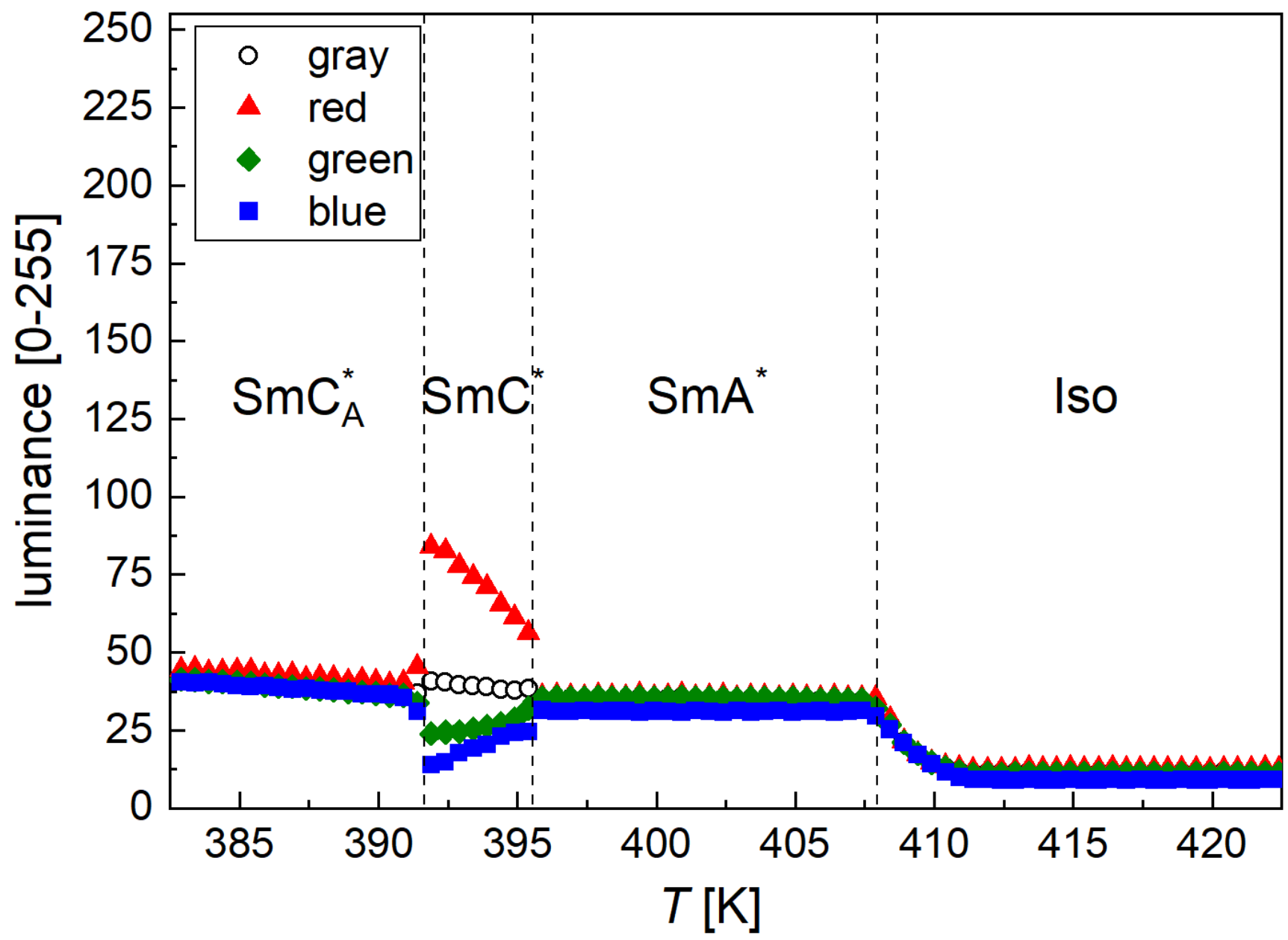

Figure S10. Representative POM textures (622 × 466 μm$^2$) of MIX5HFHH6 collected at the 2 K/min heating rate in the reflection mode as well as the red, green, blue components and weighted total luminance of each texture.

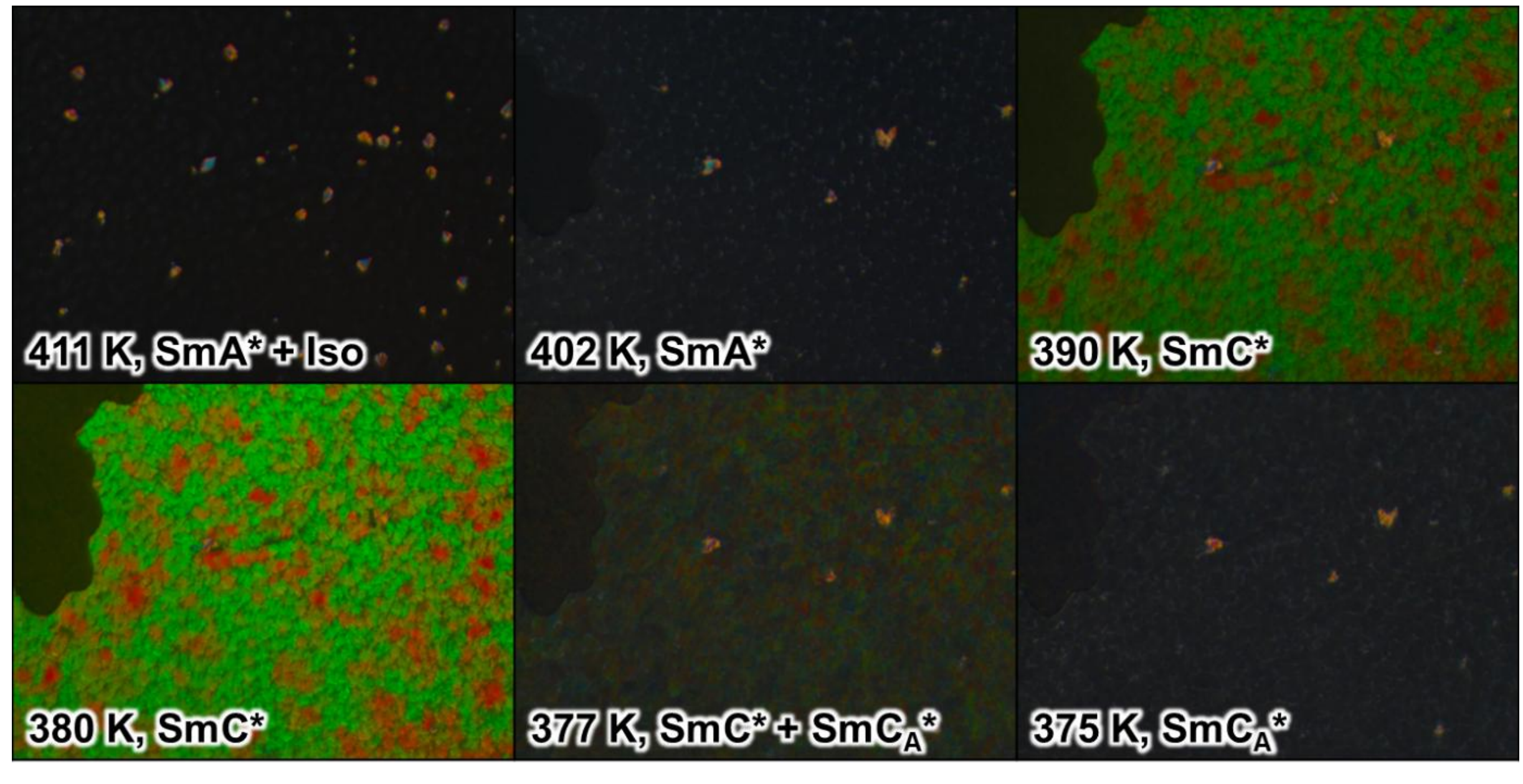

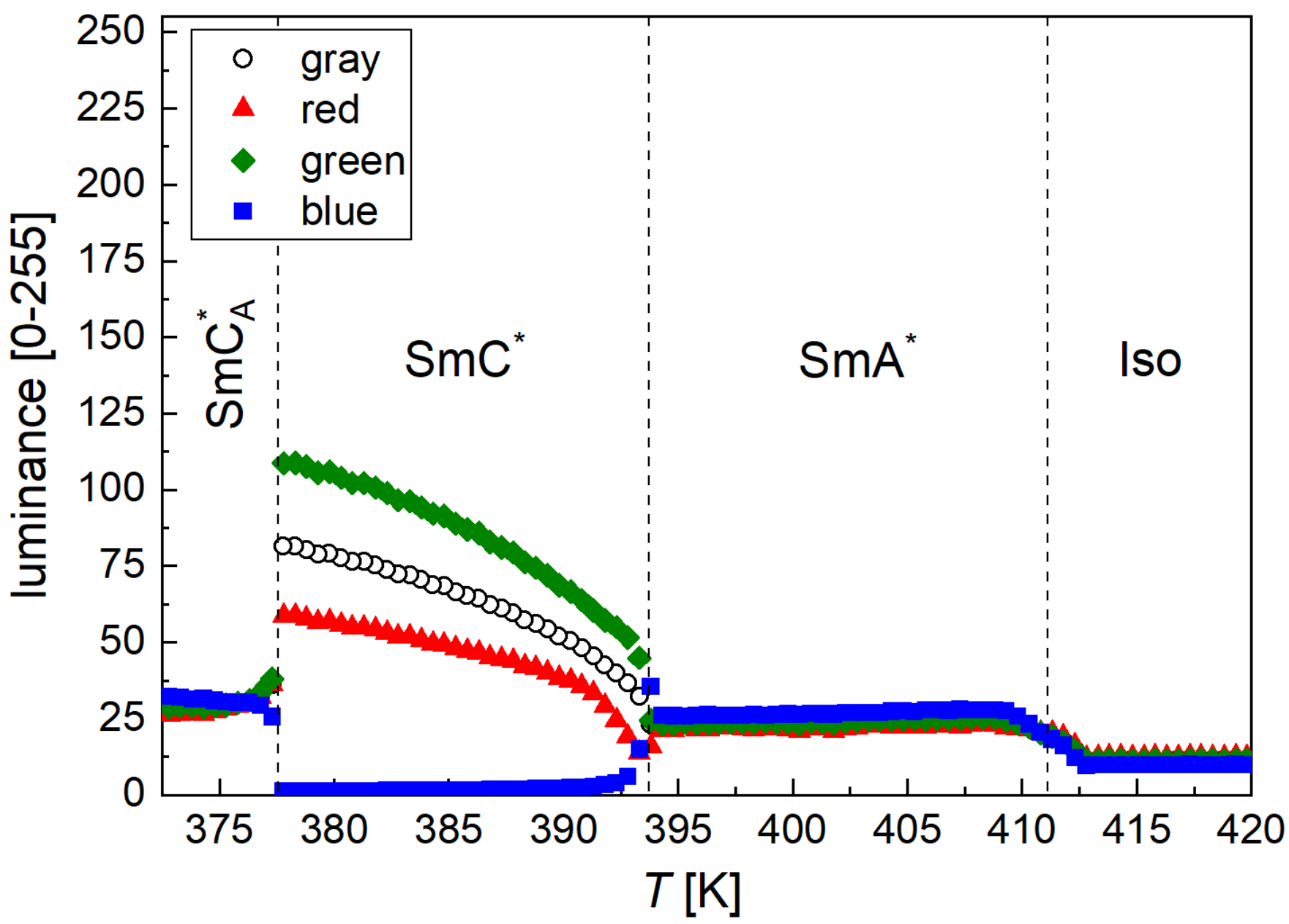

Figure S11. Representative POM textures (622 × 466 μm$^2$) of MIX6HFHH6 collected at the 2 K/min cooling rate in the reflection mode as well as the red, green, blue components and weighted total luminance of each texture.

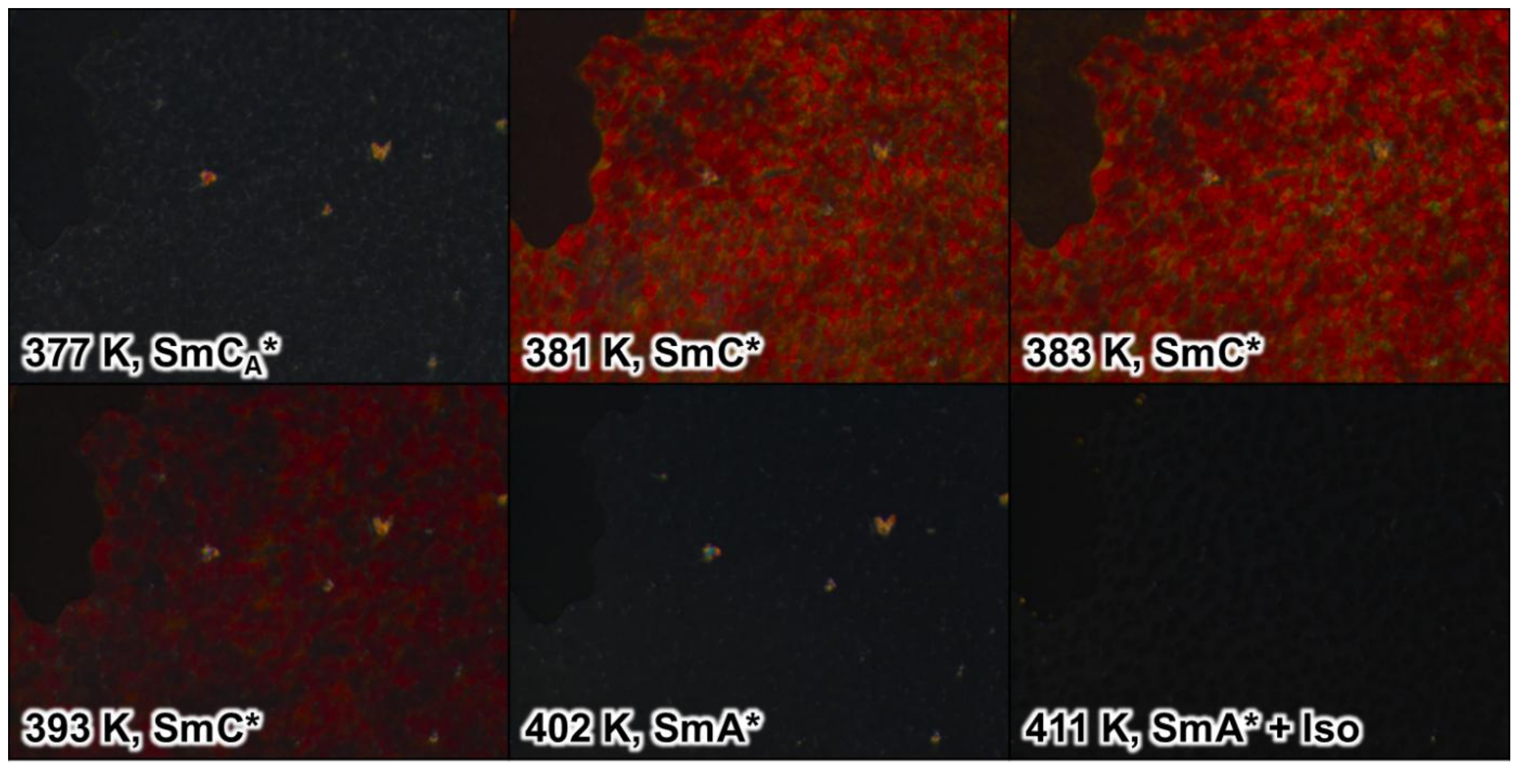

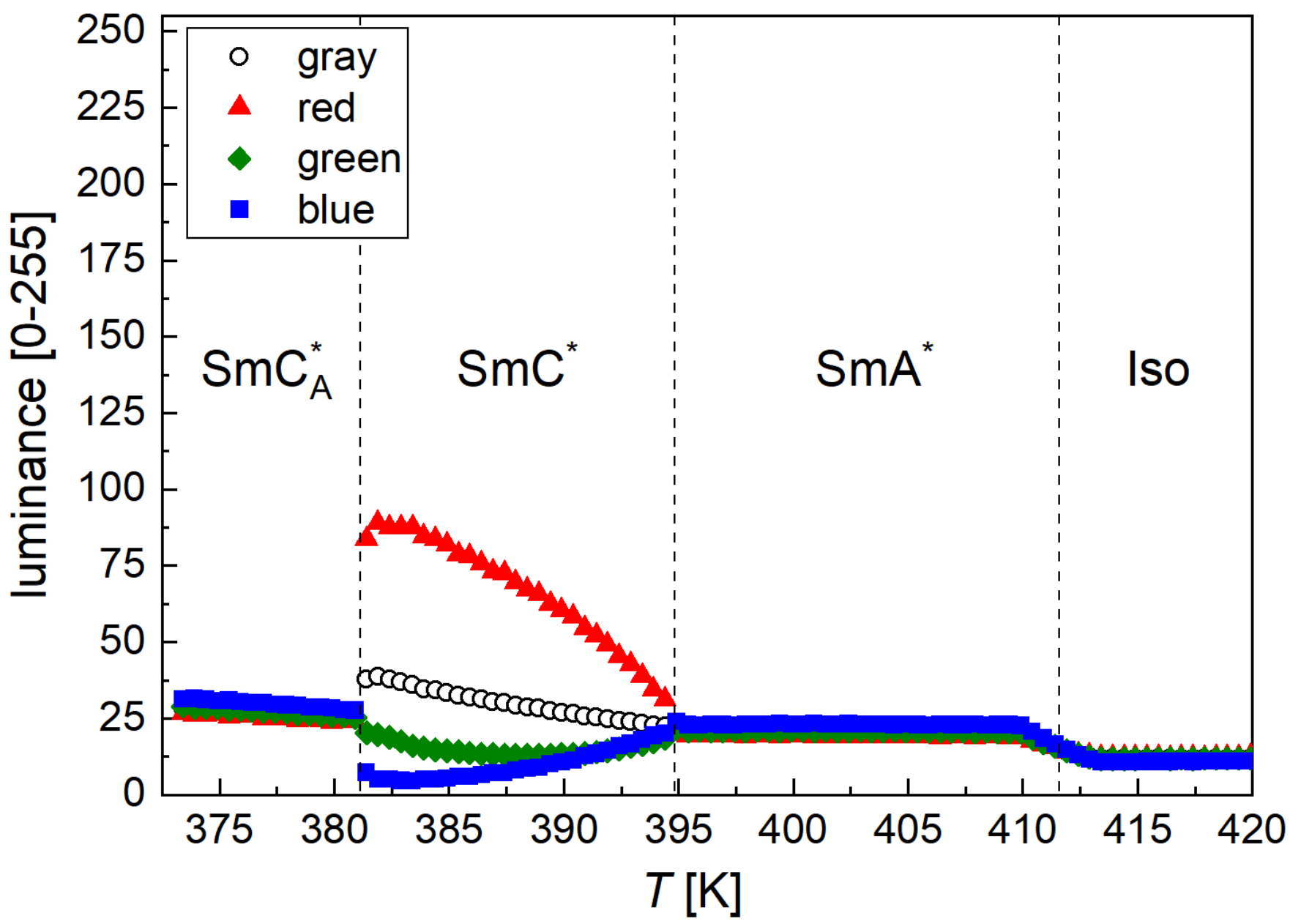

Figure S12. Representative POM textures (622 × 466 μm$^2$) of MIX6HFHH6 collected at the 2 K/min heating rate in the reflection mode as well as the red, green, blue components and weighted total luminance of each texture.

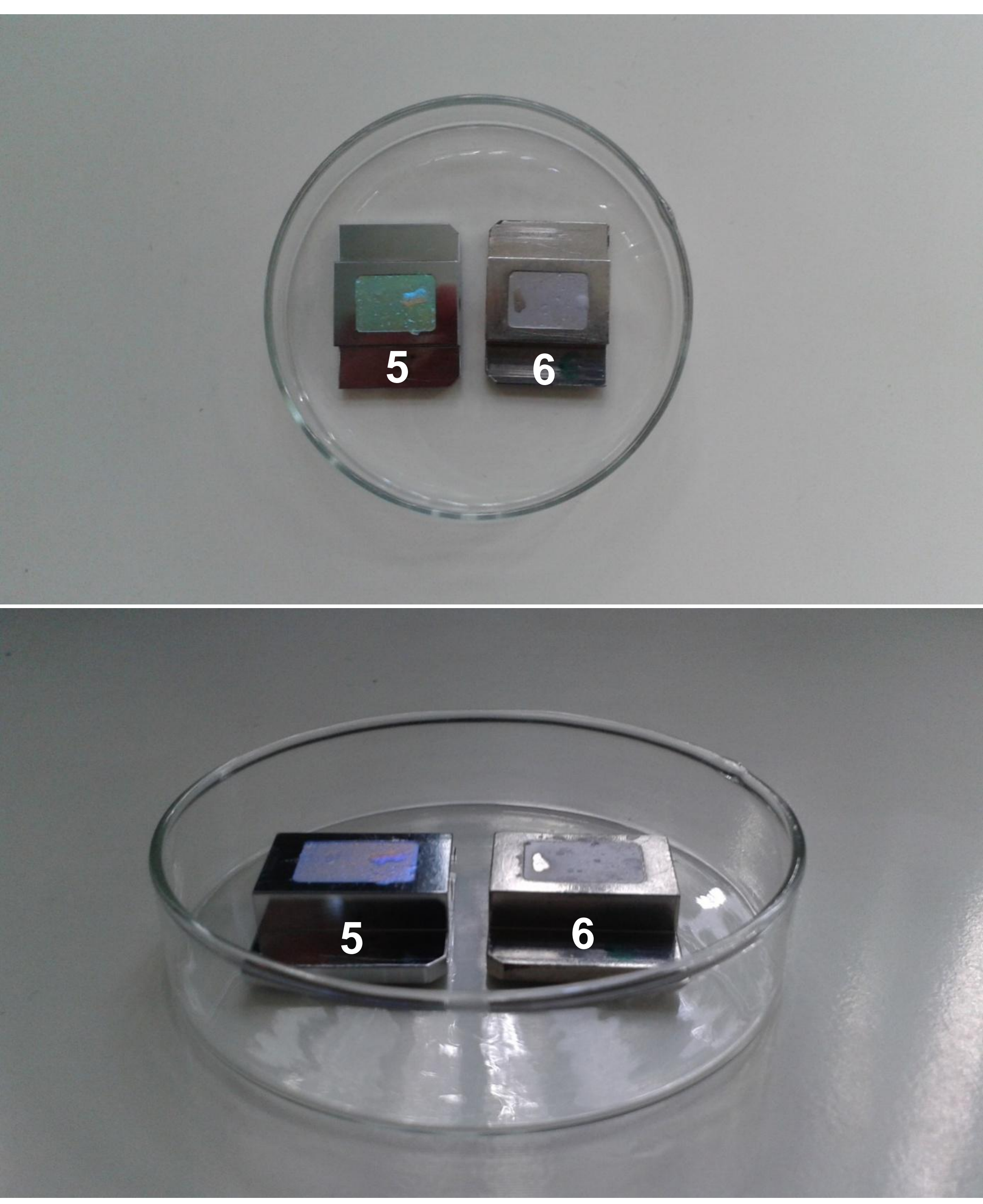

Figure S13. Homeotropically-aligned samples of MIXmHFHH6 (m = 5 on the left, m = 6 on the right) viewed from the top and from the side.

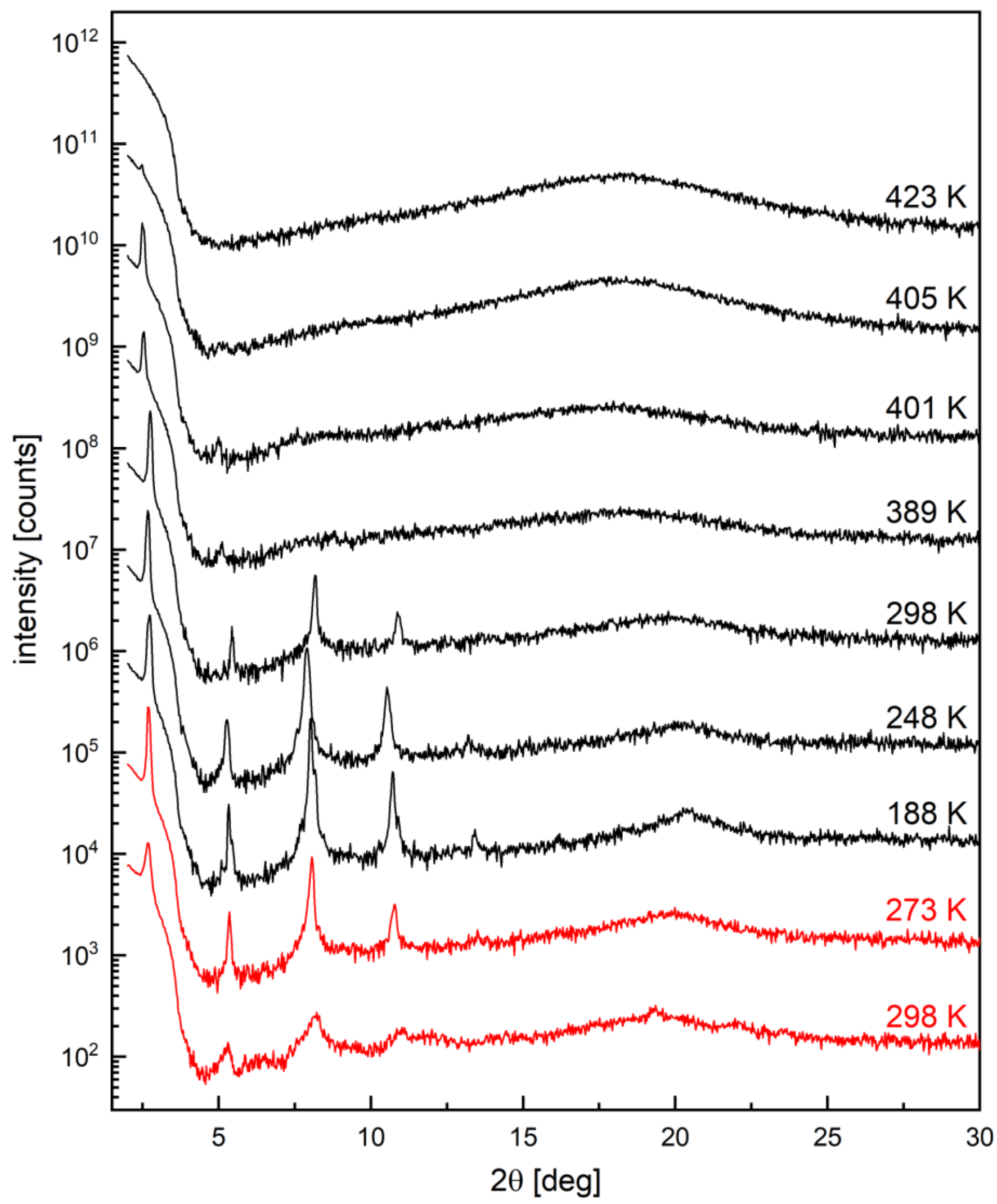

Figure S14. Selected XRD patterns of MIX5HFHH6 collected on cooling from 423 K to 188 K and on heating to 273 K and 298 K.

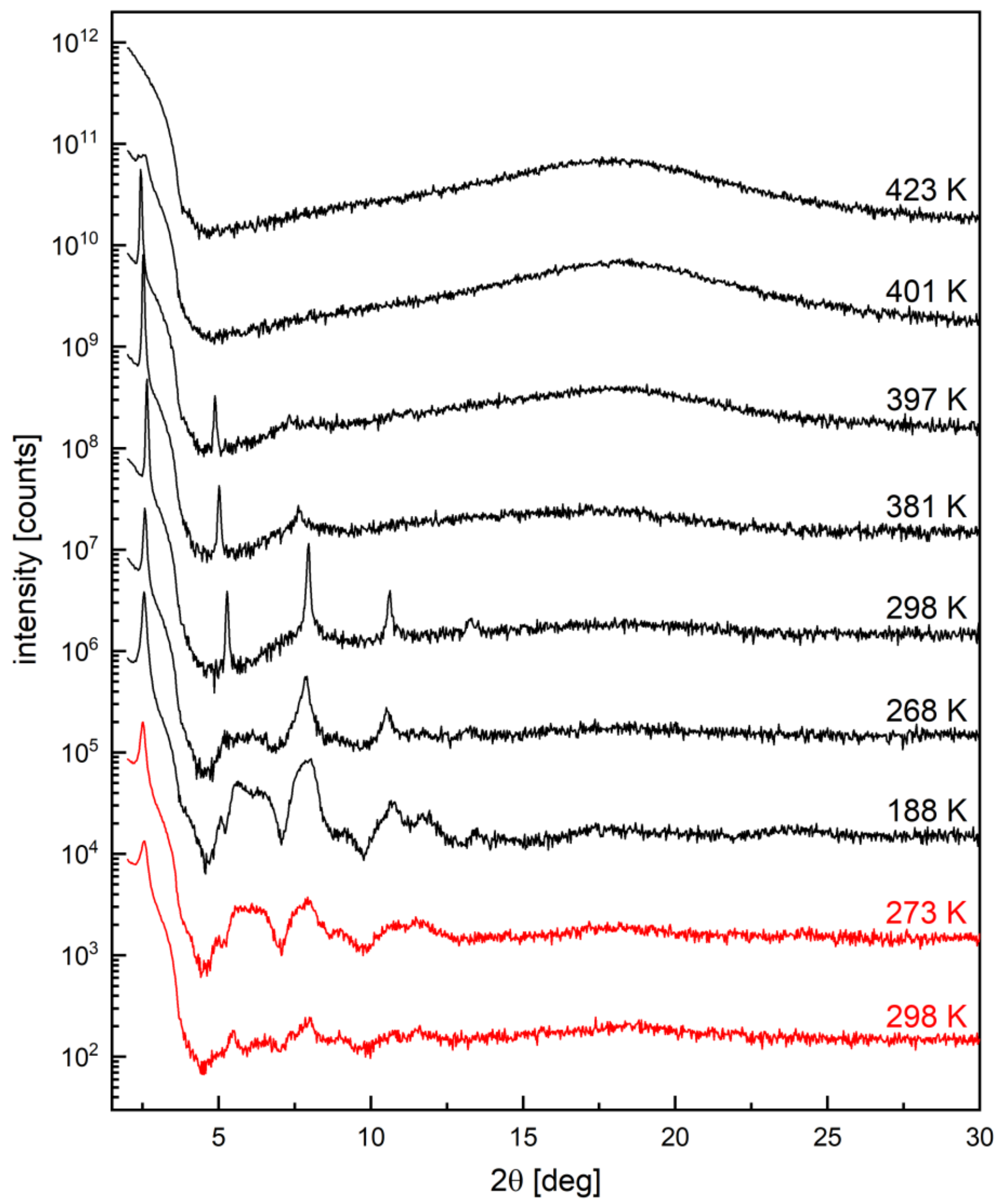

Figure S15. Selected XRD patterns of MIX6HFHH6 collected on cooling from 423 K to 188 K and on heating to 273 K and 298 K.